\documentclass{aa}  

\usepackage{graphicx}
\usepackage{array}
\usepackage{txfonts}
\usepackage{caption}
\usepackage{lipsum}
\usepackage{hyperref}
\usepackage{url}
\usepackage{natbib}
\usepackage{diagbox}
\usepackage{cuted} 
\usepackage{mathtools}
\usepackage[mathscr]{euscript}

\newcolumntype{C}{>{\centering\arraybackslash}p{0.2\textwidth}}

\usepackage{enumitem}
\usepackage{booktabs}
\bibpunct{(}{)}{;}{a}{}{,} 
\hypersetup{colorlinks=true,linkcolor=blue,citecolor=blue,urlcolor=blue}
\begin{document} 
\title{Molecular hydrogen formation on dust: The impact of gas--dust drift on 
formation efficiency}
\titlerunning{The impact of gas--dust drift on H$_2$ formation efficiency}
\authorrunning{Reissl et al.}
\author{Stefan Reissl \inst{\ref{inst1}} \and  Simon C. O. Glover \inst{\ref{inst1}} \and  Ralf S. Klessen \inst{\ref{inst1},\ref{inst2}} \and  Liam S. Morrissey \inst{\ref{inst3},\ref{inst4}} \and Mordecai-Mark Mac Low\inst{\ref{inst5}} }
\institute{Universität \centering Heidelberg, Zentrum für Astronomie, Institut für Theoretische Astrophysik, Albert-Ueberle-Str. 2, 69120 Heidelberg, Germany \label{inst1}
\and
\centering Universität Heidelberg, Interdisziplinäres Zentrum für Wissenschaftliches Rechnen, Im Neuenheimer Feld 205, 69120 Heidelberg, Germany \label{inst2}
\and
\centering Engineering and Applied Science, Memorial University, St. John’s, NL, A1B2W4, Canada \label{inst3}
\and
\centering Department of Earth \& Planetary Science, American Museum of Natural History, New York, NY 10024, USA \label{inst4}
\and
\centering Department of Astrophysics, American Museum of Natural History, New York, NY  10024, USA \label{inst5}
}

\abstract
  {Molecular hydrogen is predominantly formed on dust-grain surfaces in the  interstellar medium. Relative motion between gas and dust occurs
  in dynamically active astrophysical environments. While the dependence of H$_2$ formation efficiency on temperature and surface properties of grains is extensively studied, the impact of gas--dust drift on the formation process has received little attention.}
  {We investigate how gas--dust drift modifies H$_2$ formation on interstellar dust grains and, in particular, the competition between the
  drift-enhanced H-atom collision rate and the reduced sticking probability at increased impact energies.}
  {We develop an event-driven kinetic Monte Carlo model that follows
  individual H atoms on spherical silicate and carbonaceous grains, including
  adsorption, surface migration, and thermal desorption processes as well as Langmuir-Hinshelwood (LH) and Eley-Rideal
  (ER) reaction channels. Gas--dust drift is described by a shifted Maxwellian
  velocity distribution. We compare a constant sticking probability with a drift-dependent prescription that accounts for the increased impact energy.}
  {Gas--dust drift produces an increasingly anisotropic distribution of adsorbed H and H$_2$ formation over the grain surface. For the assumption of constant
  sticking, increasing drift generally enhances H$_2$ formation through the increased collision rate, with formation efficiencies up to $\epsilon_{\mathrm{H_2}}\simeq0.3$--0.4. When sticking depends on impact energy, however, strong drift suppresses H$_2$ formation for both materials, with the efficiency decreasing to $\epsilon_{\mathrm{H_2}}\simeq0.01$--0.03.
  Carbonaceous grains do remain efficient to higher dust temperatures  than silicate grains, though. ER reactions dominate over most of the investigated parameter
  space and become increasingly important at strong drift as the reduced
  surface population suppresses LH reactions.}
  {Gas--dust drift affects grain-surface H$_2$ formation. An enhanced collision rate does not necessarily imply a higher formation rate because increased impact energy can reduce sticking. H$_2$ formation prescriptions for dynamically active environments should therefore account for the relative gas--dust velocity and its effect on the collision rate and adsorption probability.}

  \maketitle
%
\section{Introduction}
Molecular hydrogen is the most abundant molecule in the interstellar medium (ISM) and plays a fundamental role in its chemical and thermal evolution \cite{Genzel1991,Tielens2013,Klessen2016}. It provides the starting point for much of interstellar chemistry and is an important coolant of astrophysical gas, particularly in warm star-forming environments. Consequently, while it is still debated whether H$_2$ molecules are connected to the thermal evolution of interstellar clouds and ultimately to the conditions under which stars form \citep{Glover2012MNRAS,Klessen2016SAAS} the formation of H$_2$ needs to be accounted for. Despite the apparent simplicity of the H$_2$ molecule, its formation under interstellar conditions remains a non-trivial surface-chemistry problem \citep[see][for review]{Wakelam2017}.

Direct formation of H$_2$ in the gas phase is inefficient under typical ISM conditions as it usually occurs by a three-body reaction that requires densities far above those encountered in diffuse and ordinary molecular clouds \citep[see e.g.][]{Palla1983}. Direct radiative association of two neutral H atoms is extremely inefficient \citep[see e.g.][for further details]{Forrey2016}. Catalysis by H$^-$ or H$_2^+$ ions is important in radiation-poor environments such as primordial star-forming regions, but not in the modern ISM \citep{Glover2003}. Interstellar dust grains offer an alternative formation site and simultaneously concentrate H atoms on their surfaces. Grain-surface catalytic reactions have consequently long been considered the dominant formation route for interstellar H$_2$ in galaxies with metals \citep{Gould1963ApJ,Hollenbach1971,Hollenbach1979ApJS,Purcell1979,Cazaux2004,Wakelam2017}.

Laboratory \citep[e.g.][]{Pirronello1997a,Pirronello1997b,Pirronello1999} and theoretical \citep[e.g.][]{Katz1999,Cazaux2004,Chang2005} studies subsequently demonstrated efficient H$_2$ formation on a variety of astrophysically relevant surfaces. More recent experiments indicate that H$_2$ formation on carbonaceous surfaces can remain efficient at substantially higher grain temperatures than traditionally assumed \citep{Grieco2023}.

Two principal mechanisms are commonly invoked for H$_2$ formation on dust grains. In the Langmuir-Hinshelwood (LH) mechanism, gas-phase H atoms adsorb onto the grain and subsequently migrate across the surface until two adsorbed H atoms encounter each other and react \citep{Langmuir1922,Hinshelwood1926}. At sufficiently low temperatures, diffusion may become slow, whereas at high temperatures the residence time of adsorbed H atoms becomes short. Efficient LH formation is consequently often restricted to a characteristic temperature interval \citep{Chang2005,Cazaux2004}. In contrast to LH, in the Eley-Rideal (ER) mechanism an incoming H atom from the gas phase reacts directly with an H atom already bound to the surface, without requiring the incoming atom to thermalize and diffuse across the surface \citep{Eley1941}. If the bound H atom is chemisorbed, with a high desorption energy, then this mechanism can remain important under high-temperature conditions where the residence time of weakly adsorbed atoms is too short for efficient LH molecule formation. 

The relative importance of these processes depends strongly on the microscopic properties of the grain. Interstellar refractory dust is commonly represented by silicate and carbonaceous grain populations, but the interaction of atomic
hydrogen with these materials differs substantially. Both physisorption and chemisorption may contribute to H$_2$ formation, with chemisorption providing much deeper binding sites in which H atoms can survive to higher grain
temperatures \citep{Cazaux2004}. 

Carbonaceous materials, such as amorphous carbon or regular graphite, exhibit particularly rich adsorption chemistry and have been studied extensively in connection with H$_2$ formation. Surfaces of silicate bearing minerals likewise support H$_2$ formation, although their binding energies and diffusion barriers depend strongly on composition and surface structure of the minerals \citep{Goumans2009}. Thus, even for fixed gas properties, the H$_2$ formation efficiency can depend strongly on grain composition, morphology, and temperature \citep{Katz1999,Wakelam2017}.

The modeling of H$_2$ formation therefore involves several important uncertainties. On realistic amorphous surfaces, binding energies and diffusion barriers are not single, well-defined quantities but may vary from site to site. Surface roughness and porosity modify both diffusion and residence
times, while the relative importance of surface migration changes with temperature. The sticking probability of an incident H atom introduces an additional uncertainty. This depends on its kinetic energy and, in general, on the grain temperature and surface properties. Small grains can undergo substantial stochastic temperature fluctuations following photon absorption, so that a single equilibrium dust temperature is not always an adequate description \citep{Bron2014}. 

A variety of numerical approaches have been developed to address these difficulties. Rate-equation models have been widely used because of their simplicity and their straightforward coupling to gas-phase chemical networks
\citep{Cazaux2004}. Master- and moment-equation approaches include population fluctuations statistically, while more detailed Monte Carlo (MC) techniques follow individual particles and reaction events. In particular, \citet{Chang2005} introduced a continuous-time random-walk MC treatment of H$_2$ formation, allowing surface diffusion and spatial inhomogeneity to be followed explicitly. Subsequent kinetic Monte Carlo (KMC) calculations explored surface
roughness, grain size, stochastic heating, and increasingly detailed physisorption-chemisorption networks. \citet{Iqbal2012} extended this approach over a wide temperature range for both silicate and carbonaceous grains. Such calculations have demonstrated that stochastic treatments can deviate substantially from rate equations when surface populations become small, particularly at high grain temperatures.

Relative gas--dust motion can arise through astrophysical processes such as turbulence, ambipolar diffusion, or shocks. Such conditions are particularly relevant in dynamically active environments such as protostellar and molecular outflows or turbulent interstellar clouds, as well as galactic winds, while charged grains can acquire substantial velocities relative to neutral gas. Consequently, the resulting drift speed can become comparable to, or even exceed the thermal speed of atomic hydrogen.

This relative drift could affect H$_2$ formation, as the drift speed enhances the collision rate, while the increased impact energy can reduce the probability that an atom sticks to the surface. Moreover, drift makes the incident flux anisotropic.  

Previous studies have explored in considerable detail the effects of dust temperature, gas temperature, grain size, surface morphology, adsorption energies, and stochastic heating, but have commonly treated the incident gas as an isotropic thermal distribution in the grain rest frame. However, despite the astrophysical relevance of gas--dust drift, systematic drift has generally been neglected as an explicit control parameter in microscopic simulations of H$_2$ formation. We therefore investigate H$_2$ formation in the presence of gas--dust drift using event-driven KMC simulations of individual H atoms on spherical interstellar grains.

The paper is organized as follows. In Sect.~\ref{sect:GrainModel} we outline the geometry and surface properties of the dust grains. The corrections for including drift into the collision statistics are outlined in Sect.~\ref{sect:DriftCorr}. In Sect.~\ref{sect:MigrationRates} we introduce the material-dependent mobility and evaporation rates, whereas Sect.~\ref{sect:H2model} is dedicated to an analytical solution of the formation problem. The KMC scheme used is outlined in Sect.~\ref{sect:KMCscheme}. The results are presented in Sect.~\ref{sect:Results} and discussed in Sect.~\ref{sect:Discussion}. Finally, we summarize our findings in Sect.~\ref{sect:Summary}.


\section{Dust grain model}
\label{sect:GrainModel}
We model silicate and amorphous carbon grains with radius $a$ and equidistant spacing $d_{\mathrm{s}}$ between formation sites (see Table~\ref{table:param}).
Physisorption (p) occurs when an atom is weakly bound to the grain surface by van der Waals interactions, resulting in relatively low binding energies and high surface mobility. Chemisorption (c) involves the formation of a stronger chemical bond with the surface, leading to much higher binding energies and therefore longer residence times, particularly at elevated grain temperatures.

An argument can be made that p- and c-sites are not equally distributed over the grain surface  \citep{Das2026arXiv}. However, reliable data on the exact
ratio of p-sites to c-sites remain scarce. In our model, we assume both adsorption mechanisms but assume an equal number of p- and c-sites per grain \citep{Cazaux2002}. 

We assume that adsorption sites are equally distributed over the grain surface using a Fibonacci distribution scheme, where each site is connected to its six closest
neighbors. The Fibonacci scheme places points along a spherical spiral with approximately uniform spacing, avoiding the strong clustering near the poles that occurs with regular latitude-longitude grids. This give a total number of 
\begin{equation}
    N_{\mathrm{s}} =  \frac{4\pi a^2}{d_{\mathrm{s}}^2} 
\end{equation}
formation sites and a surface area of ${ A=4\pi a^2\ N_{\mathrm{s}}^{-1} }$ per site. 

Many sticking functions for various gas temperatures ranges, materials, and surface compositions are derived in the literature \citep{Hollenbach1979ApJS,Cuppen2010,Matar2010,Chaabouni2012,Wakelam2017}. These functions agree qualitatively that sticking decreases as the gas temperature $T_{\mathrm{g}}$ and dust temperature $T_{\mathrm{d}}$ increase. 

We note that the sticking of H atoms on grain surfaces remains uncertain because it depends on complex energy exchange during collisions and the local surface structure. Existing models typically rely on simplified geometries and mono-layer materials \citep{Cuppen2010,Bonfanti2015,Shi2023,Bossion2024}. Hence, no all encompassing sticking function from first principles has been found, especially for amorphous materials and the high temperature regime.

In our modeling we therefore apply the canonical sticking function \cite{Hollenbach1979ApJS}
\begin{equation}
\mathcal{S}\left( T_{\mathrm{g}} , T_{\mathrm{d}} \right) = \left[ 1+\frac{2}{5}\left\{\left(\frac{T_{\mathrm{d}}+T_{\mathrm{g}}}{100\ \mathrm{K}}\right)^{1/2} + \left(\frac{T_{\mathrm{g}}}{100\ \mathrm{K}}\right) + \frac{1}{5}\left(\frac{T_{\mathrm{g}}}{100\ \mathrm{K}}\right)^2 \right\} \right]^{-1}
\label{eq:sticking}
\end{equation}
that at least gives the correct qualitative temperature dependence. For simplicity we apply this function for both silicate and carbon materials. Note that, strictly speaking, Equation~\eqref{eq:sticking} is valid only when the drift velocity of the grains is zero. In the next section 
we adjust this function for the case of non-zero drift velocity.

\section{Drift corrections}
\label{sect:DriftCorr}
A grain embedded in gas at rest is isotropically bombarded by gas particles in thermal motion. Here, we assume that the gas consists only of atomic hydrogen atoms with mass $m_{\rm H}$. The velocity distribution follows Maxwell-Boltzman statistics with mean thermal velocity\footnote{Note that the thermal motion of the grains themselves can be neglected here, since $m_{\rm grain} \gg m_{\rm H}$.}
\begin{equation}
v_{\rm th} = \sqrt{\frac{8k_{\rm B}T_{\rm g}}{\pi m_{\rm H}}}.
\end{equation}
where $k_{\rm B}$ is the Boltzmann constant. However, for a grain that drifts with a velocity $U$ through the gas phase the velocity distribution becomes skewed and the statistics of gas--dust collisions need to be adjusted accordingly. For an efficient evaluation we introduce the dimensionless drift parameter
\begin{equation}
s=\frac{U}{v_{\rm th}}\, .
\end{equation}
The enhancement of the rate of H atoms being injected into the simulation domain and, subsequently, colliding with the grain surface due to grain-gas drift
\begin{equation} 
R_\mathrm{coll}=  \pi a^{2}  n_{\rm H} v_{\rm th} \kappa(s)
\end{equation}
is described by the correction factor\footnote{We note that the different ways $\kappa(s)$ is presented in the literature depend on how the drift parameter $s$ is defined: either relative to the mean thermal velocity, as here, or relative to the most likely thermal velocity.}
\begin{equation}
\kappa(s) = \frac{1}{2}\exp\left(-\frac{4s^{2}}{\pi}\right) +
\left(s+\frac{\pi}{8s}\right)
\operatorname{erf}\left(\frac{2s}{\sqrt{\pi}}\right)\, .
\end{equation}
This factor is obtained by averaging the relative velocity over a Maxwellian velocity distribution shifted by the drift velocity $U$ (see \citealt{Baines1965} and \citealt{DasWeingartner2016MNRAS} for further details). 

\begin{figure}[]
	\begin{center}
	\includegraphics[width=0.5\textwidth]{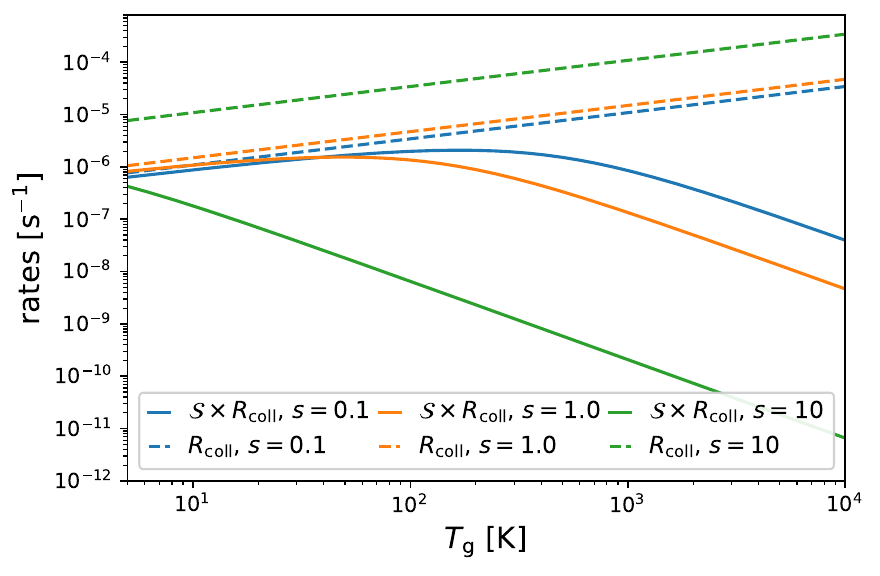}
	\end{center}
\caption{The collision rate $R_{\mathrm{coll}}$ (dashed lines) and the rate of sticking of H atoms  onto the grain surface $\mathcal{S}\times 
R_\mathrm{coll}$ (solid lines) as a function of dust temperature $T_{\mathrm{d}}$ for a drift parameter of $s=0.1$ (blue), $s=1.0$ (orange), and $s=10$ (green), respectively. The sticking function $ \mathcal{S} = \mathcal{S}( T_{\mathrm{eff}} , T_{\mathrm{d}} ) $ is evaluated including the  effective gas temperature $T_{\mathrm{eff}}$.}
\label{fig:CollRates}
\end{figure}

\begin{figure*}[]
	\begin{center}
	\includegraphics[width=1.0\textwidth]{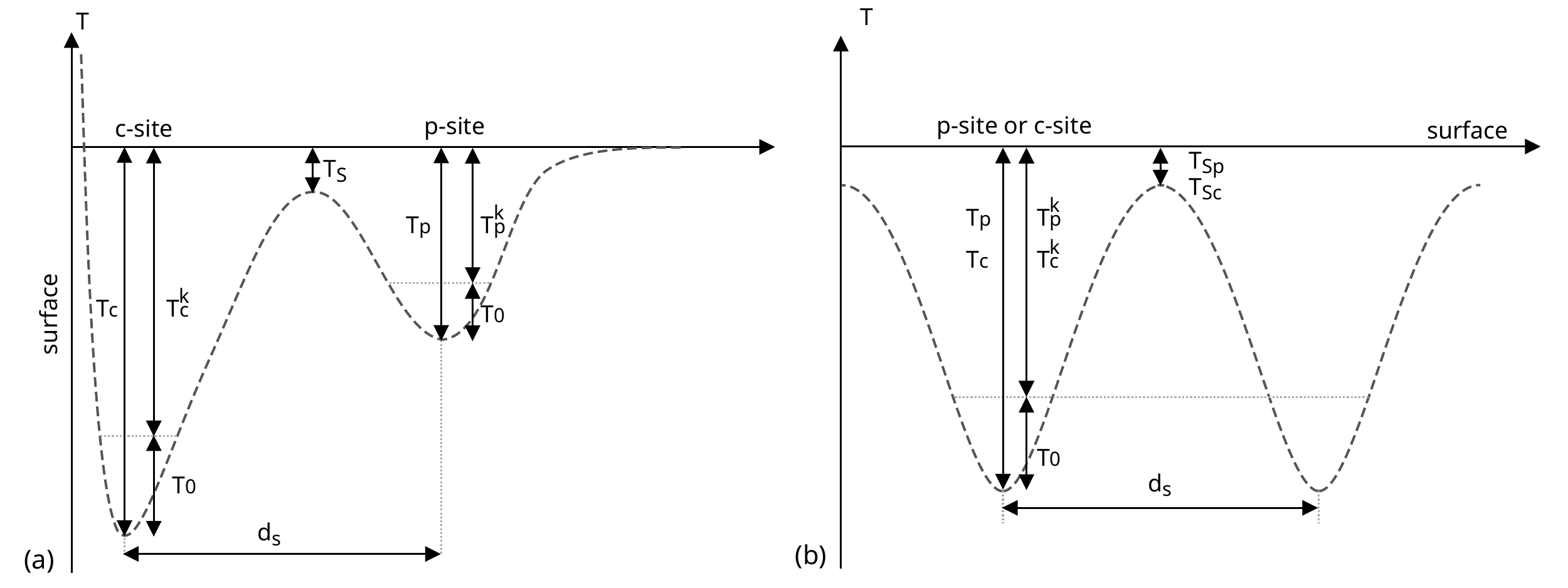}
	\end{center}
\caption{Schematic illustration of the adsorption sites and potential barriers. Energies $E$ are expressed in temperature units, $T=E/k_{\mathrm{B}}$. Panel (a) shows the physisorption (p) and chemisorption (c) sites, with binding energies $T_{\mathrm{p}}$ and $T_{\mathrm{c}}$, respectively, and the barrier $T_{\mathrm{S}}$ separating the two adsorption sites. The quantities $T_{\mathrm{p}}^{\mathrm{k}}$ and $T_{\mathrm{c}}^{\mathrm{k}}$ denote the corresponding binding energies corrected for the zero-point energy $T_0$. Panel (b) shows the lateral potential between neighboring adsorption sites. The barriers for migration between neighboring p-sites and c-sites are characterized by $T_{\mathrm{Sp}}$ and $T_{\mathrm{Sc}}$, respectively. Here, each potential minimum represents an adsorption site associated with a particular Fibonacci point on the grain surface.}
\label{fig:Potential}
\end{figure*}

Since sticking is related to gas temperature it follows that the gas impact velocity itself plays a key role. However, impact velocity is usually not covered as a separate parameter in laboratory experiments or theoretical modeling. In our modeling we evaluate the sticking function ${\cal S}$ (Eq. \ref{eq:sticking}) for an effective gas temperature
\begin{equation}
T_{\rm eff} = T_{\rm g} + 
\frac{m_{\rm H}U_{\perp}^{2}}{k_{\rm B}},
\end{equation}
accounting for the kinetic energy associated with the gas--dust drift normal to the grain surface that is defined as ${ U_{\perp}=\max\left(0,-\mathbf{U}\cdot\hat{\mathbf{n}}\right) }$, 
where $\hat{\mathbf{n}}$ is the local outward  normal of the grain surface. We emphasize that $T_{\rm eff}$ is not a unique surface temperature but is a relative quantity dependent on the orientation  of each surface element with respect to the drift velocity $\vec{U}$. Hence, the effective temperature approach describing the sticking including drift reproduces correctly the limit ${ \mathcal{S}(T_{\rm eff}, T_{\rm d}) \rightarrow \mathcal{S}(T_{\rm g}, T_{\rm d})}$ for ${ U \rightarrow  0 }$. In order to constrain the impact of temperature and drift on sticking we also consider the constant case of ${ \mathcal{S} = 0.5 }$ in our calculations. 
The interplay of collision rate $R_{\mathrm{coll}}$ and the 
sticking function $\mathcal{S}( T_{\mathrm{eff}} , T_{\mathrm{d}} )$ evaluated with an effective temperature $T_{\mathrm{eff}}$ is shown in Fig.~\ref{fig:CollRates} for different drift parameters $s$.

\section{Migration and evaporation rates}
\label{sect:MigrationRates}
In our model H atoms coming from the gas phase are deposited on the grain surface by entering an empty p-site. From there they can move to a neighboring p-site by means of thermal hopping or quantum tunneling. Alternatively, the H atom can enter the corresponding c-site. In either case the H atom has to overcome a relative energy barrier of depth B$_{\mathrm{X}}$.\footnote{We use sub-indices ${ \mathrm{X} \in \{\mathrm{p},\mathrm{c}\} }$ and ${ \mathrm{Y} \in \{\mathrm{p},\mathrm{c}} \}$ as place holders for physisorption p and chemisorption c in order to introduce the equations for the permutations pp, cc, pc, and cp most efficiently.} The depths and relative positions of the adsorption sites and potential barriers are illustrated in Fig.~\ref{fig:Potential}.

To evaluate the transmission probability, we use the analytical solution of the one-dimensional Schrödinger equation for a rectangular potential barrier \citep[see, e.g.,][]{Messiah1961,Cazaux2004}. In the general case considered here, the potential energies on the two sides of the barrier are not necessarily equal, as is particularly relevant for transitions between p-sites and c-sites (see Fig.~\ref{fig:Potential}). The resulting transmission coefficient for
a transition from a site of type X to a site of type Y is
\begin{equation}
 \Lambda_{\mathrm{XY}} =4  \begin{cases}  \cfrac{ \sqrt{ \frac{T - B_{\mathrm{XY}}}{T} }  }{ \left( 1 + \sqrt{ \frac{T - B_{\mathrm{XY}}}{T} }  \right)^2  + \cfrac{ B_{\mathrm{X}} B_{\mathrm{Y}}   \sinh^2{\left(d_{\mathrm{s}} / \lambda_{\mathrm{D}} \right) }    }{ (B_{\mathrm{X}} - T)T }   } & \mbox{if } T \leq B_{\mathrm{X}}  \vspace{0.25cm} \\   \cfrac{ \sqrt{ \frac{T - B_{\mathrm{XY}}}{T} } }{ 
 \left( 1 + \sqrt{ \frac{T - B_{\mathrm{XY}}}{T} }  \right)^2  + \cfrac{ B_{\mathrm{X}} B_{\mathrm{Y}}   \sin^2{\left(d_{\mathrm{s}} / \lambda_{\mathrm{D}} \right) }   }{ (B_{\mathrm{X}} - T)T }   } & \mbox{otherwise}\end{cases}.
\end{equation}
Here, $T$ denotes the energy of an adsorbed H atom expressed in temperature units, $T=E/k_{\rm B}$, while $B_{\rm X}$ and $B_{\rm Y}$ characterize the barrier heights relative to the potential minima of the initial and final sites, respectively, and $B_{\rm XY}$ accounts for the difference between the potentials on the two sides of the barrier. The corresponding values for the different transitions are given in Table~\ref{tab:BXY}. For p-p and c-c transitions, the potential minima on the two sides are equal and $B_{\rm XY}=0$, such that the expression reduces to the standard result for a rectangular barrier with identical potentials on either side. For p-c and c--p transitions, the two potential minima differ and the more general form above is required.

For $T\leq B_{\rm X}$, passage through the barrier occurs by quantum tunneling, whereas for $T>B_{\rm X}$ the expression describes above-barrier transmission. The quantity
\begin{equation} 
\lambda_{\rm D}
=
\hbar\left(2m_{\rm H}k_{\rm B}|T-B_{\rm X}|\right)^{-1/2}
\end{equation}
sets the characteristic length scale of the wave function within the barrier.

The probability of an adsorbed H atom moving to another site can be directly related to the dust temperature $T_{\mathrm{d}}$ by integrating over all possible energy sites \citep{Cazaux2002} to yield
\begin{equation}
 P_{\mathrm{XY}} = \frac{1}{T_{\mathrm{d}}} \int_{0}^{\infty}  \Lambda_{\mathrm{XY}} \exp\left( -\frac{T}{T_{\mathrm{d}}} \right) \mathrm{d}T.
\end{equation}
The rates for migration between adsorption sites are defined by
\begin{equation}
  \alpha_{\mathrm{XY}} =  \nu_{\mathrm{X}} P_{\mathrm{XY}},
\end{equation}
where 
\begin{equation}
  \nu_{\mathrm{X}} =  \frac{1}{\pi} \sqrt{  \frac{2k_{\mathrm{B}} T_{\mathrm{X}} }{m_{\mathrm{H}} d_{\mathrm{s}}^2}  }
\end{equation}
is the characteristic vibrational frequency of the adsorbed H atom and $T_{\mathrm{X}}$ is the potential depth (see Fig.~\ref{fig:Potential}). In addition to H$_2$ formation by means of the ER and LH mechanisms, which we assume always results in the loss of the molecule from the surface, the adsorbed H atom may also escape the grain surface by means of thermal desorption. The corresponding desorption rate may be evaluated via 
\begin{equation}
  \beta_{\mathrm{X}} =  \nu_{\mathrm{X}} \exp\left( -\frac{T_{\mathrm{X}}^{\mathrm{k}}}{T_{\mathrm{d}}} \right).
\end{equation}
Furthermore, it needs to be considered that an adsorbed H atom does not lie in the absolute minimum of the adsorption potential but at its zero-point energy $T_0 = h \nu_\mathrm{X} / (2 k_\mathrm{B})$. Hence, we correct each potential depth to
\begin{equation}
    T_{\mathrm{X}}^{\mathrm{k}} = T_{\mathrm{X}} - T_{\mathrm{0}}. 
\end{equation}

\begin{figure*}[ht!]
	\begin{center}
	\includegraphics[width=1.02\textwidth]{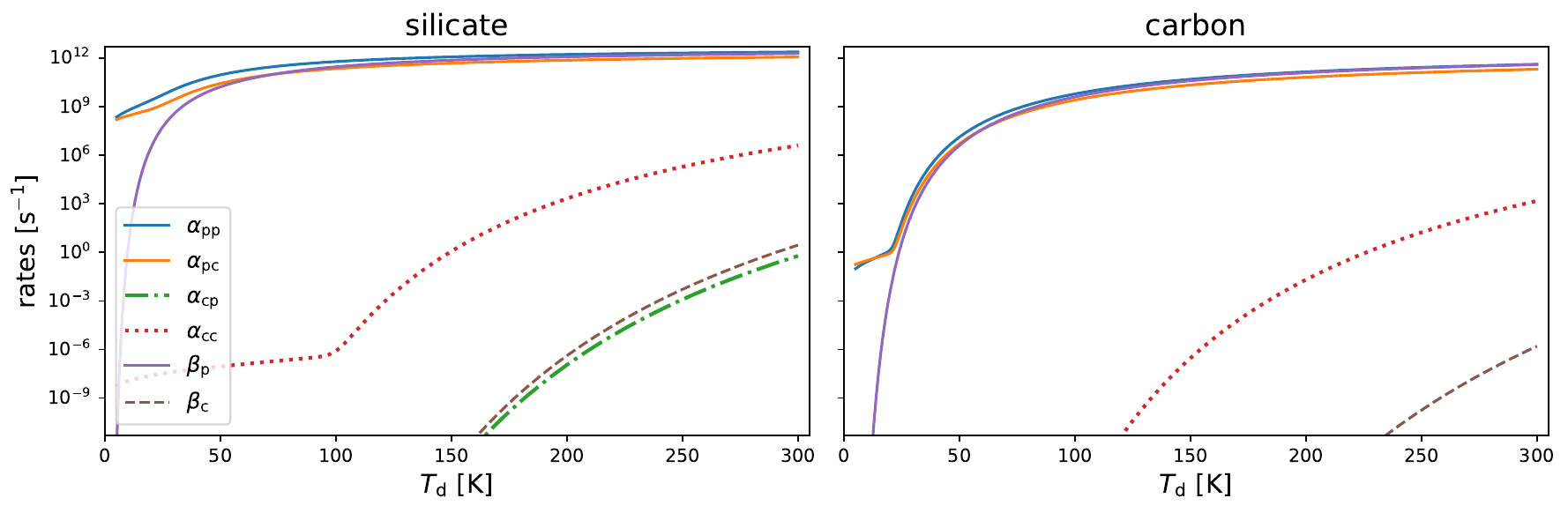}
	\end{center}
\caption{Surface mobility and desorption rates of adsorbed H atoms as a function of dust temperature $T_{\mathrm{d}}$ for silicate (left panel) and carbon grains (right panel). Shown are the pp and cc hopping rates $\alpha_{\mathrm{pp}}$ (blue solid) and $\alpha_{\mathrm{cc}}$ (red dotted), the pc and cp transition rates $\alpha_{\mathrm{pc}}$ (orange solid) and $\alpha_{\mathrm{cp}}$ (green dash-dotted), and the desorption rates from p-sites and c-sites, $\beta_{\mathrm{p}}$ (purple solid) and $\beta_{\mathrm{c}}$ (brown dashed), respectively.}
\label{fig:MigrationRates}
\end{figure*}

The p- and c-site potentials, $T_{\mathrm{p}}$ and $T_{\mathrm{c}}$, respectively, as well as the height of the barriers $T_{\mathrm{S}}$, $T_{\mathrm{Sp}}$, and $T_{\mathrm{Sc}}$ are shown in Fig.~\ref{fig:Potential} and their values are listed in Table~\ref{table:param}, whereas all permutations of  values of $B_{\mathrm{XY}}$ considered are given in Table~\ref{tab:BXY}. Here, we follow the approach of \cite{Cazaux2004} to estimate the barriers by taking $T_{\mathrm{Sp}}=T_{\mathrm{S}}$ and $T_{\mathrm{Sc}}=T_{\mathrm{c}}/2$.
The migration rates $\alpha_{\mathrm{XY}}$ and evaporation rates $\beta_{\mathrm{X}}$ are shown in Fig.~\ref{fig:MigrationRates} for the range of considered dust temperatures $T_{\mathrm{d}}$. 


\begin{table}
\centering
\caption{Adsorption site properties} 
 \begin{tabular}{c|cccc} 
 \hline
  material & 
  $T_{\mathrm{p}}\ [\mathrm{K}]$ &  $T_{\mathrm{c}}\ [\mathrm{K}]$ &  $T_{\mathrm{S}}\ [\mathrm{K}]$ & $d_{\mathrm{s}}\ [\mathrm{nm}]$ \\
 \hline
 silicate & $400$ & $10\ 000$ & $100$ & $0.17$   \\
 carbon & $780$ & $14\ 000$ & $100$ & $0.3$    \\

 \hline
\end{tabular}
\label{table:param}
\tablefoot{Potential depths $T_{\mathrm{p}}$ and $T_{\mathrm{c}}$ of the p-sites and c-sites, respectively, and the average separation $d_{\mathrm{s}}$ between
adsorption sites. The values are adopted from \cite{Cazaux2004,Chang2005,Perets2006,Cuppen2005,Bron2014} to represent typical silicate and amorphous carbon surfaces. For simplicity, we approximate $T_{\mathrm{Sp}}=T_{\mathrm{S}}$ and $T_{\mathrm{Sc}}=T_{\mathrm{c}}/2$ and assume a uniform spatial separation $d_{\mathrm{s}}$ between all neighboring potential minima \citep{Iqbal2012}.}
\end{table}

\begin{table}
\centering
\caption{Effective potential barriers $B$ considered for transitions between
physisorption (p) and chemisorption (c) sites.}
\begin{tabular}{c|ccc}
\hline
 & $B_{\mathrm{X}}\ [\mathrm{K}]$
 & $B_{\mathrm{Y}}\ [\mathrm{K}]$
 & $B_{\mathrm{XY}}\ [\mathrm{K}]$ \\
\hline
$\mathrm{pp}$
& $T_{\mathrm{p}}^{\mathrm{k}}-T_{\mathrm{Sp}}$
& $T_{\mathrm{p}}^{\mathrm{k}}-T_{\mathrm{Sp}}$
& $0$ \\

$\mathrm{cc}$
& $T_{\mathrm{c}}^{\mathrm{k}}-T_{\mathrm{Sc}}$
& $T_{\mathrm{c}}^{\mathrm{k}}-T_{\mathrm{Sc}}$
& $0$ \\

$\mathrm{pc}$
& $T_{\mathrm{p}}^{\mathrm{k}}-T_{\mathrm{S}}$
& $T_{\mathrm{c}}^{\mathrm{k}}-T_{\mathrm{S}}$
& $T_{\mathrm{p}}^{\mathrm{k}}-T_{\mathrm{c}}^{\mathrm{k}}$ \\

$\mathrm{cp}$
& $T_{\mathrm{c}}^{\mathrm{k}}-T_{\mathrm{S}}$
& $T_{\mathrm{p}}^{\mathrm{k}}-T_{\mathrm{S}}$
& $T_{\mathrm{p}}^{\mathrm{k}}-T_{\mathrm{c}}^{\mathrm{k}}$ \\

\hline
\end{tabular}
\label{tab:BXY}
\tablefoot{Effective potential barriers $B$ considered for transitions between
physisorption (p) and chemisorption (c) sites.}
\end{table}

\section{Analytical model of H$_2$ formation} 
\label{sect:H2model}
Given the dust model and the rates presented in the previous sections, the
time evolution of the number of H atoms occupying physisorption sites,
$N_{\rm p}$, is given by
\begin{align}
\frac{{\rm d}N_{\rm p}}{{\rm d}t}
={}&
\mathcal{S}R_{\rm coll}
(1-\theta_{\rm p})(1-\theta_{\rm c})
-R_{\rm coll}\theta_{\rm p}(1-\theta_{\rm c})
-\frac{1}{2}R_{\rm coll}\theta_{\rm p}\theta_{\rm c}
\nonumber\\
&-\alpha_{\rm pc}N_{\rm p}
+\alpha_{\rm cp}N_{\rm c}(1-2\theta_{\rm p})
-2\alpha_{\rm pp}N_{\rm p}\theta_{\rm p}
-\beta_{\rm p}N_{\rm p},
\label{eq:dNp_dt}
\end{align}
while the number of H atoms occupying chemisorption sites, $N_{\rm c}$,
evolves according to
\begin{align}
\frac{{\rm d}N_{\rm c}}{{\rm d}t}
={}&
\alpha_{\rm pc}N_{\rm p}(1-2\theta_{\rm c})
-\alpha_{\rm cp}N_{\rm c}
-2\alpha_{\rm cc}N_{\rm c}\theta_{\rm c}
-\beta_{\rm c}N_{\rm c}
\nonumber\\
&-R_{\rm coll}(1-\theta_{\rm p})\theta_{\rm c}
-\frac{1}{2}R_{\rm coll}\theta_{\rm p}\theta_{\rm c},
\label{eq:dNc_dt}
\end{align}
where we substitute  $\theta_{\rm p}=N_{\rm p}/N_{\rm s}$ and
$\theta_{\rm c}=N_{\rm c}/N_{\rm s}$.

\cite{Bron2014} argue that H atoms from the gas phase may enter c-sites
directly and suggest an additional sticking function for chemisorption.
However, we follow the approach of \cite{Kolasinski2008}, in which incoming
H atoms first enter a p-site. The individual terms in the
rate equations can then be interpreted as follows.

The term
$\mathcal{S}R_{\rm coll}(1-\theta_{\rm p})(1-\theta_{\rm c})$
describes adsorption of an incoming gas-phase H atom when both the p- and
c-sites of the corresponding surface site are empty. If the p-site is
occupied while the c-stite is empty, an incoming H atom reacts with the
physisorbed H atom through the ER mechanism, giving the contribution
$-R_{\rm coll}\theta_{\rm p}(1-\theta_{\rm c})$ to
Eq.~\eqref{eq:dNp_dt}. Similarly, if the c-site is occupied while the
p-site is empty, the ER reaction removes a chemisorbed H atom, giving
$-R_{\rm coll}(1-\theta_{\rm p})\theta_{\rm c}$ in
Eq.~\eqref{eq:dNc_dt}. If both sites are occupied, the incoming H atom
reacts with either the physisorbed or chemisorbed H atom with equal
probability. The corresponding expectation values therefore give
$-\frac{1}{2}R_{\rm coll}\theta_{\rm p}\theta_{\rm c}$ in each population
equation.

Transitions from physisorption to chemisorption occur at a total rate
$\alpha_{\rm pc}N_{\rm p}$. Every p$\rightarrow$c transition removes one
atom from the physisorbed population, giving
$-\alpha_{\rm pc}N_{\rm p}$ in Eq.~\eqref{eq:dNp_dt}. The change in the
chemisorbed population depends on the occupation of the destination
c-site. A transition onto an empty c-site increases $N_{\rm c}$ by one,
whereas a transition onto an occupied c-site produces H$_2$ and removes
the previously chemisorbed H atom. The mean contribution is therefore
${
\alpha_{\rm pc}N_{\rm p}
\left[(1-\theta_{\rm c})-\theta_{\rm c}\right]
=
\alpha_{\rm pc}N_{\rm p}(1-2\theta_{\rm c})}$.

The reverse c$\rightarrow$p transition occurs at a total rate
$\alpha_{\rm cp}N_{\rm c}$. Every such transition removes one atom from
the chemisorbed population, giving
$-\alpha_{\rm cp}N_{\rm c}$ in Eq.~\eqref{eq:dNc_dt}. For the
physisorbed population, a transition onto an empty p-site increases
$N_{\rm p}$ by one, whereas a transition onto an occupied p-site produces
H$_2$ and removes the previously physisorbed H atom. Hence, the mean
contribution to Eq.~\eqref{eq:dNp_dt} is
${
\alpha_{\rm cp}N_{\rm c}
\left[(1-\theta_{\rm p})-\theta_{\rm p}\right]
=
\alpha_{\rm cp}N_{\rm c}(1-2\theta_{\rm p})}$.

Surface diffusion provides two additional H$_2$ formation channels. A
p-H atom hopping onto an occupied neighboring p-site removes two
physisorbed H atoms, giving
$-2\alpha_{\rm pp}N_{\rm p}\theta_{\rm p}$, while a c-H atom hopping onto
an occupied neighboring c-site removes two chemisorbed H atoms, giving
$-2\alpha_{\rm cc}N_{\rm c}\theta_{\rm c}$. Finally,
$-\beta_{\rm p}N_{\rm p}$ and $-\beta_{\rm c}N_{\rm c}$ describe thermal
desorption from p- and c-sites, respectively.

If stochastic processes such as temperature fluctuations, ultraviolet photodesorption, and cosmic-ray induced desorption are ignored, we can derive the equilibrium solutions $N_{\mathrm{p,eq}}$ and $N_{\mathrm{c,eq}}$ for physisorption and chemisorption, respectively, by numerically evaluating ${ \mathrm{d} N_{\mathrm{p}} / \mathrm{d} t = 0 }$ and ${ \mathrm{d} N_{\mathrm{c}} / \mathrm{d} t = 0 }$. The corresponding H$_2$ formation rate  is given by

\begin{equation}
R_{\rm H_2}
= \alpha_{\rm pp}
N_{\rm p}\theta_{\rm p} + \alpha_{\rm cc}
N_{\rm c}\theta_{\rm c}
+ \alpha_{\rm pc}
N_{\rm p}\theta_{\rm c}
+ \alpha_{\rm cp}
N_{\rm c}\theta_{\rm p}
+
R_{\rm coll} \left(\theta_{\rm p}
+ \theta_{\rm c}
- \theta_{\rm p}\theta_{\rm c}\right)
.
\label{eq:RH2}
\end{equation}
The formation efficiency
\begin{equation}
\epsilon_{\mathrm{H_2}} = \frac{2 R_{\mathrm{H_2}}}{R_{\rm coll}}
\end{equation}
quantifies the ratio of the number of actually formed H$_2$ molecules to H atoms colliding with the grain. Note that with this definition, our $\epsilon_{\mathrm{H_{2}}}$ indirectly includes the effects of sticking \citep[e.g.][]{Iqbal2012,Bron2014}. Care should be taken when comparing this to results of other studies in the literature ignoring sticking \citep[e.g.][]{Cazaux2004,Zhao22} that report the recombination efficiency, i.e.\ the fraction of H atoms that are already stuck to the surface that react to form H$_{2}$, since the latter is equivalent to $\epsilon_{\mathrm{H_{2}}} / \mathcal{S}$ in our notation.

\section{Kinetic Monte Carlo H$_2$ formation scheme}
\label{sect:KMCscheme}
Naturally, the problem of where exactly on the grain surface H$_2$ formation
takes place can hardly be covered by analytical solutions. Hence, we use a
KMC simulation scheme\footnote{At the level of individual adsorbed H atoms,
the surface motion corresponds to a continuous-time random walk (CTRW), a
terminology often used in the literature. In this work, however, we use the
equally descriptive term KMC to emphasize that the full system evolves
through competing stochastic events, including gas arrival, surface
diffusion, desorption, transitions between p- and c-sites, and surface reactions.} to study H$_2$ formation on the grain
surface including gas--dust drift.

The collisions and subsequent evolution of adsorbed H atoms are simulated
using exponentially distributed waiting times and an event queue, in which
individual events are inserted in chronological order and the earliest
scheduled event is executed next
\citep[see, e.g.,][for further details]{Cuppen2005,Iqbal2012}.

Following \cite{Gillespie1976}, the collision events are sampled according to
\begin{equation}
\Delta t_{\mathrm{coll}}=-\frac{\ln z}{R_{\mathrm{coll}}},
\end{equation}
where $z$ is a newly generated random number uniformly sampled from
$z\in(0,1)$. After each time interval $\Delta t_{\mathrm{coll}}$, a new H
atom is injected into the simulation domain and colliding with the grain surface.

Each injected H atom is assigned a velocity drawn from a shifted
Maxwell--Boltzmann distribution,
\begin{equation}
\mathbf{v}=\mathbf{U}
+\sqrt{\frac{k_{\rm B}T_{\rm g}}{m_{\rm H}}}\,\boldsymbol{\xi},
\end{equation}
where the components of $\boldsymbol{\xi}$ are independently drawn from a
standard normal distribution. The thermal velocity components therefore
have a standard deviation $\sqrt{k_{\rm B}T_{\rm g}/m_{\rm H}}$, while
$\mathbf{U}$ accounts for the systematic gas--dust drift. The collision geometry is sampled consistently with the incident velocity distribution, and the resulting impact position on the grain surface is assigned to the nearest Fibonacci point.

If the corresponding p- or c-site is occupied, the
incoming H atom can react directly with the adsorbed H atom to form H$_2$.
If both sites are occupied simultaneously, the incoming atom reacts with
either the physisorbed or chemisorbed H atom with equal probability. If both
sites are empty, the incoming atom occupies the p-site with probability
$\mathcal{S}$, that is, if $z<\mathcal{S}$ for a newly drawn uniform random
number $z$.

For each adsorbed atom, the next surface event is scheduled in the event
queue. For a physisorbed atom, the waiting time is
\begin{equation}
\Delta t_{\mathrm{i,p}}=
-\frac{\ln z}
{\alpha_{\mathrm{pp}}+\alpha_{\mathrm{pc}}+\beta_{\mathrm{p}}}.
\end{equation}
The event is then selected according to the relative rates
$\alpha_{\mathrm{pp}}$, $\alpha_{\mathrm{pc}}$, and $\beta_{\mathrm{p}}$.
The atom therefore either hops to a neighboring p-site, transitions to the
c-site at the same surface site, or thermally desorbs. A p--p hop onto an
occupied p-site produces H$_2$, while a p--c transition onto an occupied
c-site likewise produces H$_2$.

For a chemisorbed atom, the waiting time is similarly given by
\begin{equation}
\Delta t_{\mathrm{i,c}}=
-\frac{\ln z}
{\alpha_{\mathrm{cc}}+\alpha_{\mathrm{cp}}+\beta_{\mathrm{c}}}.
\end{equation}
The event is selected according to the relative rates
$\alpha_{\mathrm{cc}}$, $\alpha_{\mathrm{cp}}$, and $\beta_{\mathrm{c}}$.
The atom therefore either hops to a neighboring c-site, transitions to the
p-site at the same surface site, or thermally desorbs. A c--c hop onto an
occupied c-site produces H$_2$, while a c--p transition onto an occupied
p-site also produces H$_2$.

Thus, H$_2$ formation occurs through either surface reactions or direct
gas--surface reactions. Surface formation occurs when a p-H or c-H atom
hops onto an occupied site of the same type, or when an atom transitions
between p- and c-site while the destination site is occupied. Direct
ER formation occurs when an incoming gas-phase H atom encounters
an occupied adsorption site. All considered events are listed in
Table~\ref{tab:mc_processes}.

To account for the numerical fluctuations inherent to KMC simulations,
equilibrium is tested using time-averaged surface populations over
consecutive time intervals. For each interval, the mean populations
$\langle N_{\rm p}\rangle$ and $\langle N_{\rm c}\rangle$ are calculated
and compared with those of the preceding interval. Equilibrium is considered
to be reached when the changes between consecutive intervals fall below a
specified threshold. Unless stated otherwise, we use the fiducial input
parameters listed in Table~\ref{tab:fiducial}.

\begin{table}
\centering
\caption{Interaction processes}
\label{tab:mc_processes}
\begin{tabular}{l|ccc}
\hline
Process
& $\Delta N_{\rm p}$
& $\Delta N_{\rm c}$
& $\Delta N_{\rm H_2}$ \\
\hline

Gas H sticks to empty p-site
& $+1$ & $0$ & $0$ \\

Gas H hits occupied p-site
& $-1$ & $0$ & $+1$ \\

Gas H hits occupied c-site
& $0$ & $-1$ & $+1$ \\

Gas H hits occupied p- and c-site$^{a}$
& $-\frac{1}{2}$ & $-\frac{1}{2}$ & $+1$ \\

p $\rightarrow$ p, target p-site empty
& $0$ & $0$ & $0$ \\

p $\rightarrow$ p, target p-site occupied
& $-2$ & $0$ & $+1$ \\

p $\rightarrow$ c, target c-site empty
& $-1$ & $+1$ & $0$ \\

p $\rightarrow$ c, target c-site occupied
& $-1$ & $-1$ & $+1$ \\

c $\rightarrow$ p, target p-site empty
& $+1$ & $-1$ & $0$ \\

c $\rightarrow$ p, target p-site occupied
& $-1$ & $-1$ & $+1$ \\

c $\rightarrow$ c, target c-site empty
& $0$ & $0$ & $0$ \\

c $\rightarrow$ c, target c-site occupied
& $0$ & $-2$ & $+1$ \\

p-H thermal desorption
& $-1$ & $0$ & $0$ \\

c-H thermal desorption
& $0$ & $-1$ & $0$ \\

\hline
\end{tabular}
\tablefoot{KMC processes considered for H atoms interacting with the grain surface. The occupation of c-sites does not affect diffusion between p-sites, and vice versa. Transitions between p- and c-sites occur at the same surface location, whereas p--p and c--c diffusion occurs between neighboring sites. $^{a}$If both adsorption sites are occupied, the incoming H atom reacts with either the p-H or c-H atom with equal probability. The listed changes in
$N_{\rm p}$ and $N_{\rm c}$ therefore represent expectation values.}
\end{table}

\begin{figure}[]
	\begin{center}
	\includegraphics[width=0.5\textwidth]{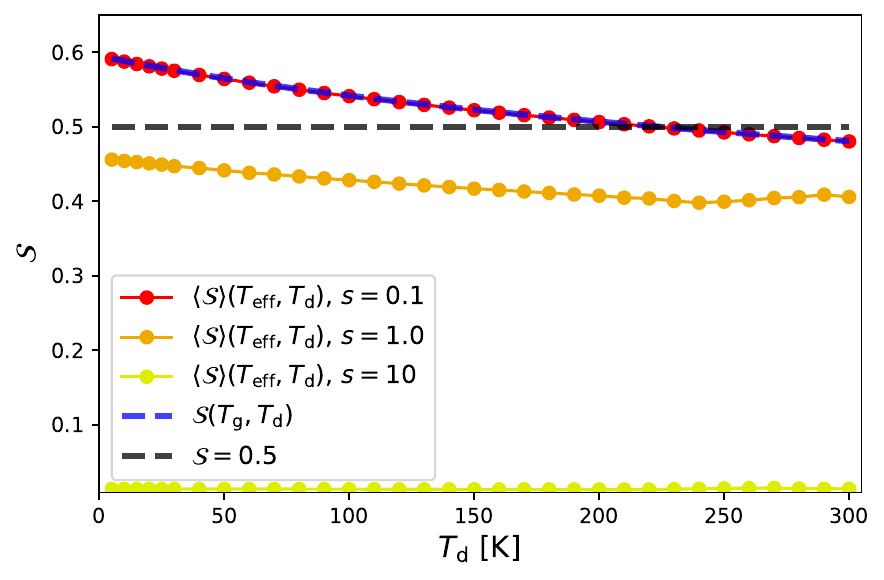}
	\end{center}
\caption{The time-averaged sticking function $\langle \mathcal{S}  \rangle (T_{\mathrm{eff}}, T_{\mathrm{d}})$ plotted as a function of the dust temperature $T_{\rm d}$. Values are shown for a gas temperature $T_{\rm g} = 100$~K and for drift parameters of $s=0.1$ (red), $s=1.0$ (orange), and $s=10$ (yellow). For comparison we also show the 
canonical sticking function  $\mathcal{S}(T_{\mathrm{g}}, T_{\mathrm{d}})$ for the same gas temperature (blue) and a constant sticking probability of $\mathcal{S}=0.5$ (black dashed). }
\label{fig:sticking}
\end{figure}

In Fig.~\ref{fig:sticking} we show the sticking function $\mathcal{S}( T_{\mathrm{g}} , T_{\mathrm{d}} )$ as a function of the dust temperature $T_{\mathrm{d}}$, computed assuming our fiducial gas temperature of 100~K. For comparison, we also show the averaged sticking $\langle \mathcal{S}( T_{\mathrm{eff}} , T_{\mathrm{d}} )  \rangle$ that we find within KMC simulations with drift parameters of $s=0.1$, $s=1.0$, and $s=10$, respectively. Naturally, for $s=0.1$ both $\mathcal{S}( T_{\mathrm{g}} , T_{\mathrm{d}} )$ and $\langle \mathcal{S}( T_{\mathrm{eff}} , T_{\mathrm{d}} )  \rangle$ are nearly identical. On the other hand, when $s$ is large, the average sticking $\langle \mathcal{S}( T_{\mathrm{eff}} , T_{\mathrm{d}} )  \rangle$ approaches zero for the entire considered range of $T_{\mathrm{d}}$. Our conclusion that the H$_2$ formation rate is dramatically reduced for a high gas--dust drift does rely on the assumption that our treatment of the sticking function $\mathcal{S}( T_{\mathrm{eff}} , T_{\mathrm{d}} )$ with an effective temperature approach provides a phenomenologically accurate description of sticking on real dust grains.

\begin{figure}[]
	\begin{center}
 \includegraphics[width=0.5\textwidth]{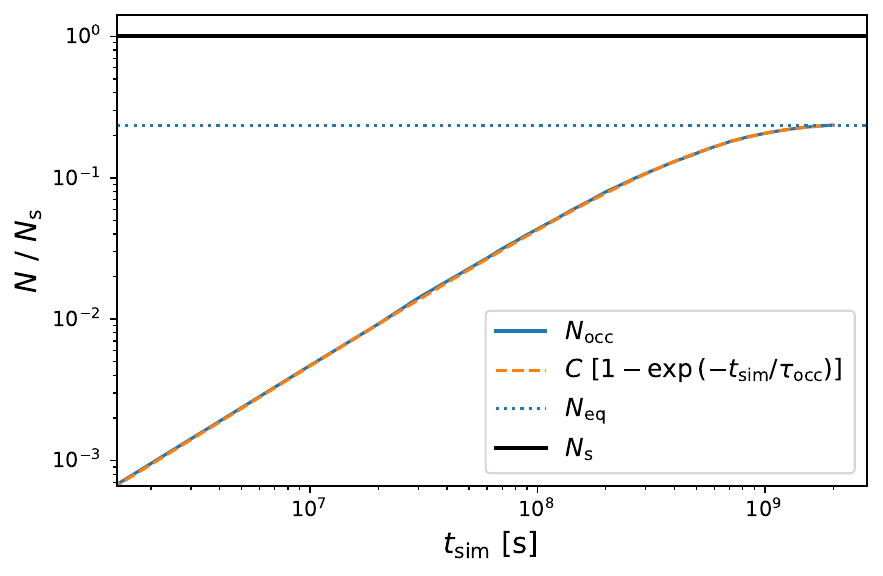}
	\end{center}
\caption{An example of the time evolution of the fraction of the sites on the surface of a carbon grain that are occupied, $N_{\rm occ}$ (solid blue line) compared to the total number of available sites $N_{\rm S}$. The simulation uses the fiducial values listed in Table~\ref{tab:fiducial} but with $s=0$. For comparison, we also show the analytical equilibrium solution $N_{\mathrm{eq}}$ (dotted blue line). The expression ${C\ [1-\exp{(-t_{\rm sim}/\tau_{\rm occ})}] }$, indicated by the orange dashed line, is a fit to $N_{\rm occ}$ used to determine the characteristic timescale $\tau_{\mathrm{occ}}$.}
\label{fig:occupation}
\end{figure}

In Fig.~\ref{fig:occupation} we show an example of the time evolution of the total number of occupied sites. Since the p-sites have multiple channels to depopulate in our model, p-sites are sparsely populated. Hence, we only show the evolution with simulation time $t_{\mathrm{sim}}$ of the total occupation number ${ N_{\mathrm{occ}} = N_{\mathrm{p}} + N_{\mathrm{c}} }$, normalized by the total number of sites  on the grain surface $N_{\mathrm{s}}$. For comparison, we also indicate the equilibrium solution  ${ N_{\mathrm{eq}} = N_{\mathrm{p,eq}} + N_{\mathrm{c,eq}} }$ (see Sect.~\ref{sect:H2model}), again normalized here by $N_{\rm s}$. We see that $N_{\rm occ} / N_{\rm s}$ in the KMC simulation stabilizes at the equilibrium value,  
demonstrating the reliability of our KMC implementation. Moreover, we see that in this example, the equilibrium fraction of occupied sites is around $20\ \%$, i.e.\ the grain surface in equilibrium never becomes fully covered by H atoms. Additionally, we fit the relation ${ N_{\mathrm{occ}} = C\ [1-\exp(-t_{\mathrm{sim}} / \tau_{\mathrm{occ}}) ]}$ to each simulation result, where $A$ is a scaling factor, in order to determine a characteristic timescale $\tau_{\mathrm{occ}}$ for the occupation process. In the example shown here, $\tau_{\mathrm{occ}} = 2.6 \times 10^8\ \mathrm{s}$. We note that smaller grains will have a lower drift velocity than larger ones, as they generally have shorter aerodynamic stopping times \citep[e.g.][]{Laibe2012}. 

Large grains with radii of order $1\,\mu\mathrm{m}$ or extending into the 
mm-regime, as well as  simulations at substantially higher gas densities, are currently computationally demanding because of the rapidly increasing number of surface sites and stochastic events. In principle, such limitations could be alleviated using hierarchical fractional-step approximations or $\tau$-leaping schemes \citep{Arampatzis2012,2008Vlachos,Che2024}. However, in the present work we employ the event-driven KMC approach commonly used for grain-surface H$_2$ formation \citep{Cazaux2004,Iqbal2012}. We therefore restrict our calculations to grain radii $a\lesssim10\ \mathrm{nm}$ and to the density
range considered in this work. 

\begin{table}
\centering
\caption{Input parameters of the KMC simulations.}
\begin{tabular}{c|cc} 
\hline
parameter & symbol & value \\
\hline
grain radius       & $a$              & $10\ \mathrm{nm}$ \\
dust temperature   & $T_{\mathrm{d}}$ & $20\ \mathrm{K}$ \\
gas temperature    & $T_{\mathrm{g}}$ & $100\ \mathrm{K}$ \\
gas number density & $n_{\mathrm{H}}$ & $3\times10^7\ \mathrm{m}^{-3}$ \\
drift parameter    & $s$              & $0.1$ \\
\hline
\end{tabular}
\label{tab:fiducial}
\end{table}


%



\begin{figure*}[ht!]
	\begin{center}
	\includegraphics[width=1.0\textwidth]{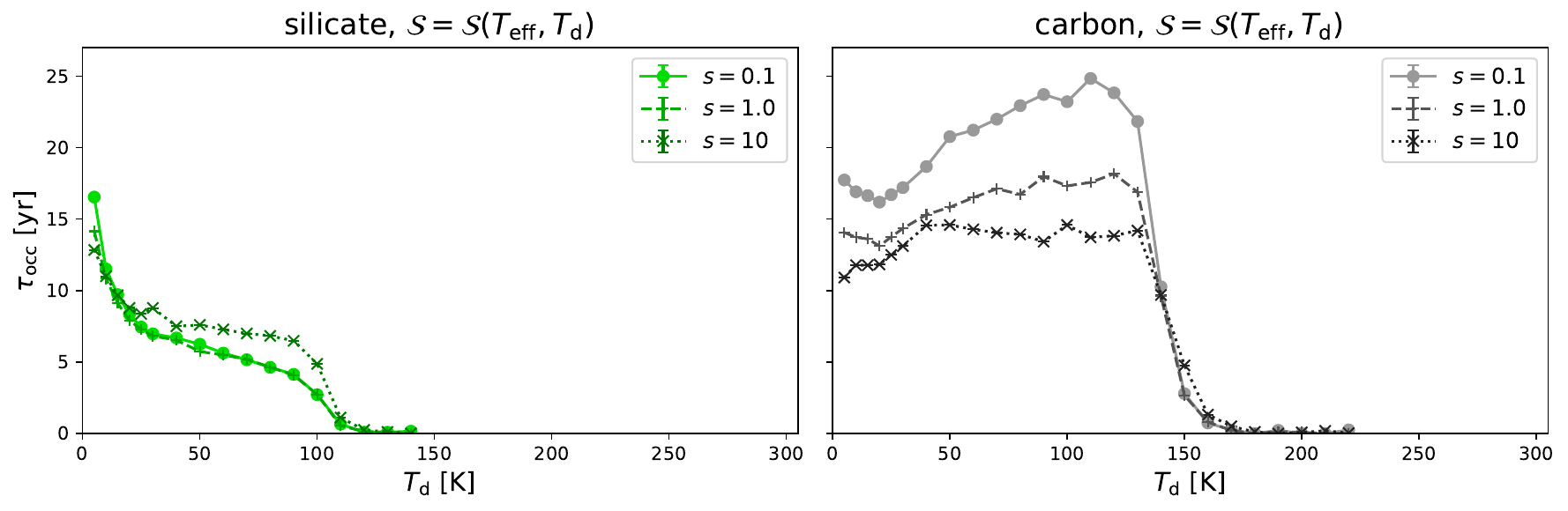}
	\end{center}
\caption{The characteristic occupation timescale $\tau_{\mathrm{occ}}$ plotted as a function of dust temperature $T_{\mathrm{d}}$ for drift parameters of $s=0.1$ (solid lines), $s=1.0$ (dashed lines), $s=10$ (dotted lines) using the temperature-dependent sticking function $ \mathcal{S} = \mathcal{S}( T_{\mathrm{eff}} , T_{\mathrm{d}} ) $ given by Eq.~\eqref{eq:sticking}, where the effective gas temperature $T_{\mathrm{eff}}$ accounts for the influence of the drift velocity. The left panel (green lines) is for silicate and the right panel (gray lines) for carbon grains. Data points are the average 
over all 32  KMC parameter sets simulated. The standard deviation of the variations among runs can be smaller than the plot symbols.}
\label{fig:TauVar}
\end{figure*}

\begin{figure*}[ht!]
	\begin{center}
	\includegraphics[width=1.0\textwidth]{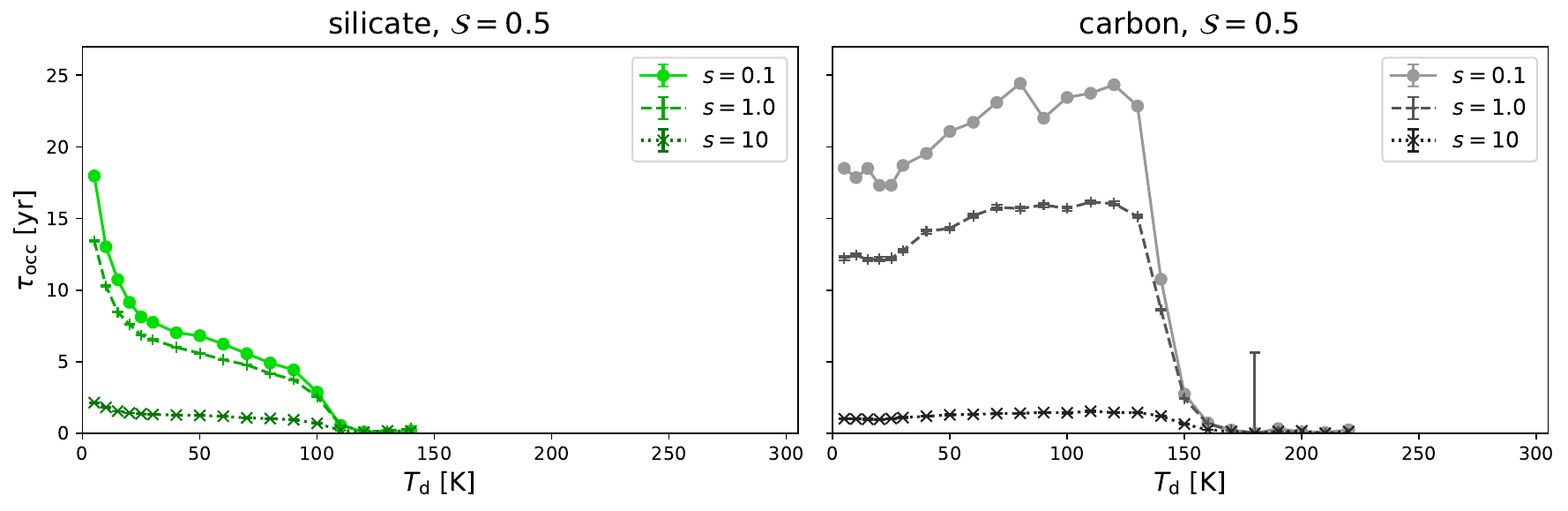}
	\end{center}
\caption{As Fig.~\ref{fig:TauVar}, but for a constant sticking function of $\mathcal{S} = 0.5$.}
\label{fig:TauConst}
\end{figure*}

\begin{figure}[ht!]
	\begin{center}
	\includegraphics[width=0.45\textwidth]{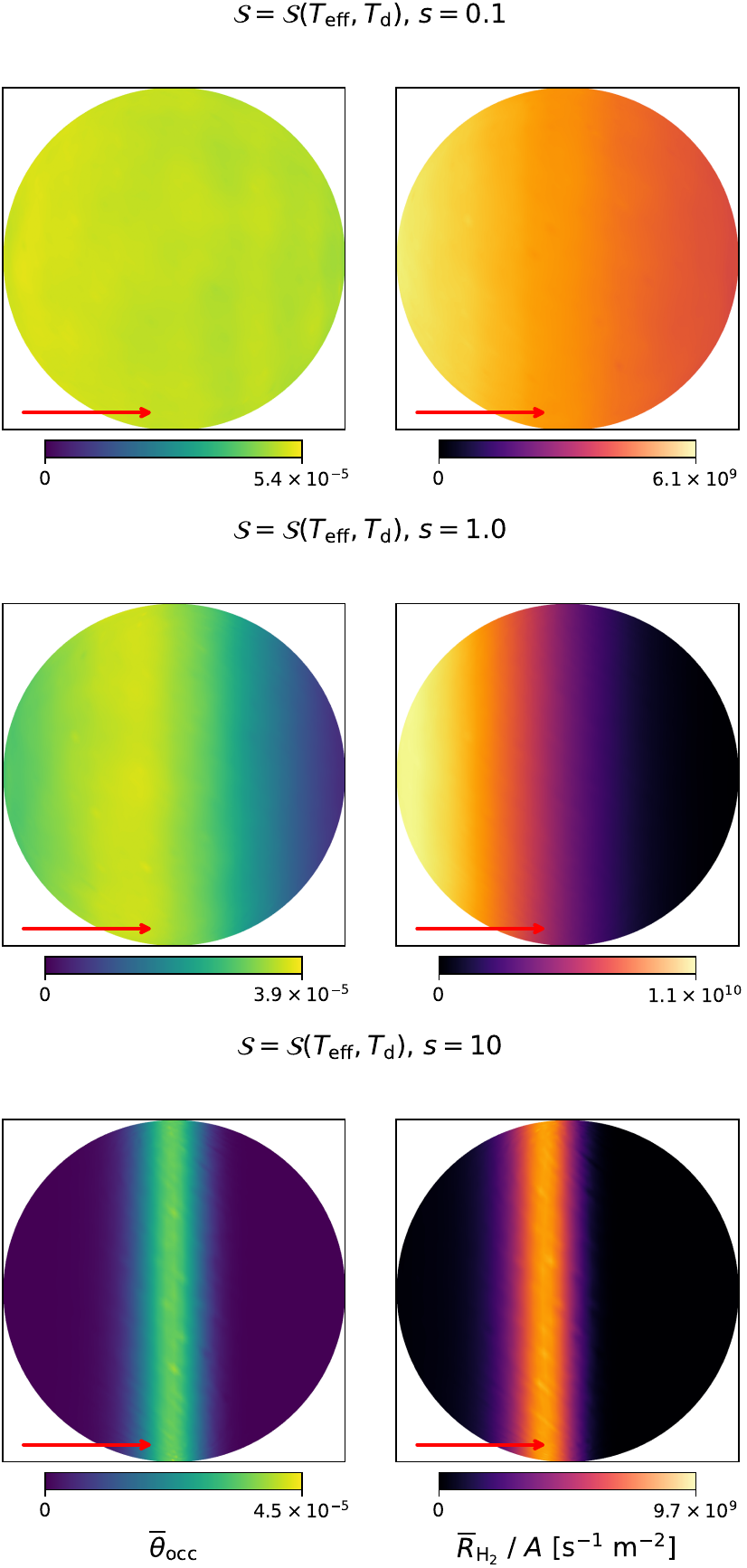}
	\end{center}
\caption{ The average surface occupation fraction ${ \overline{\theta}_{\mathrm{occ}} = N_{\mathrm{occ}} / N_{\mathrm{s}} }$ (left column) and formation rate $\overline{r}_{\mathrm{H_2}}$ per surface area $A$ (right column) on the 3D surface of a carbon grain with drift parameters of $s=0.1$ (top row), $s=1.0$ (middle row), and $s=10$ (bottom row), using a temperature dependent sticking function $\mathcal{S} =  \mathcal{S}( T_{\mathrm{eff}} , T_{\mathrm{d}} )  $. The red arrows indicate the direction of the drift velocity $\vec{U}$.  }
\label{fig:SurfaceVar}
\end{figure}

\begin{figure}[ht!]
	\begin{center}
	\includegraphics[width=0.45\textwidth]{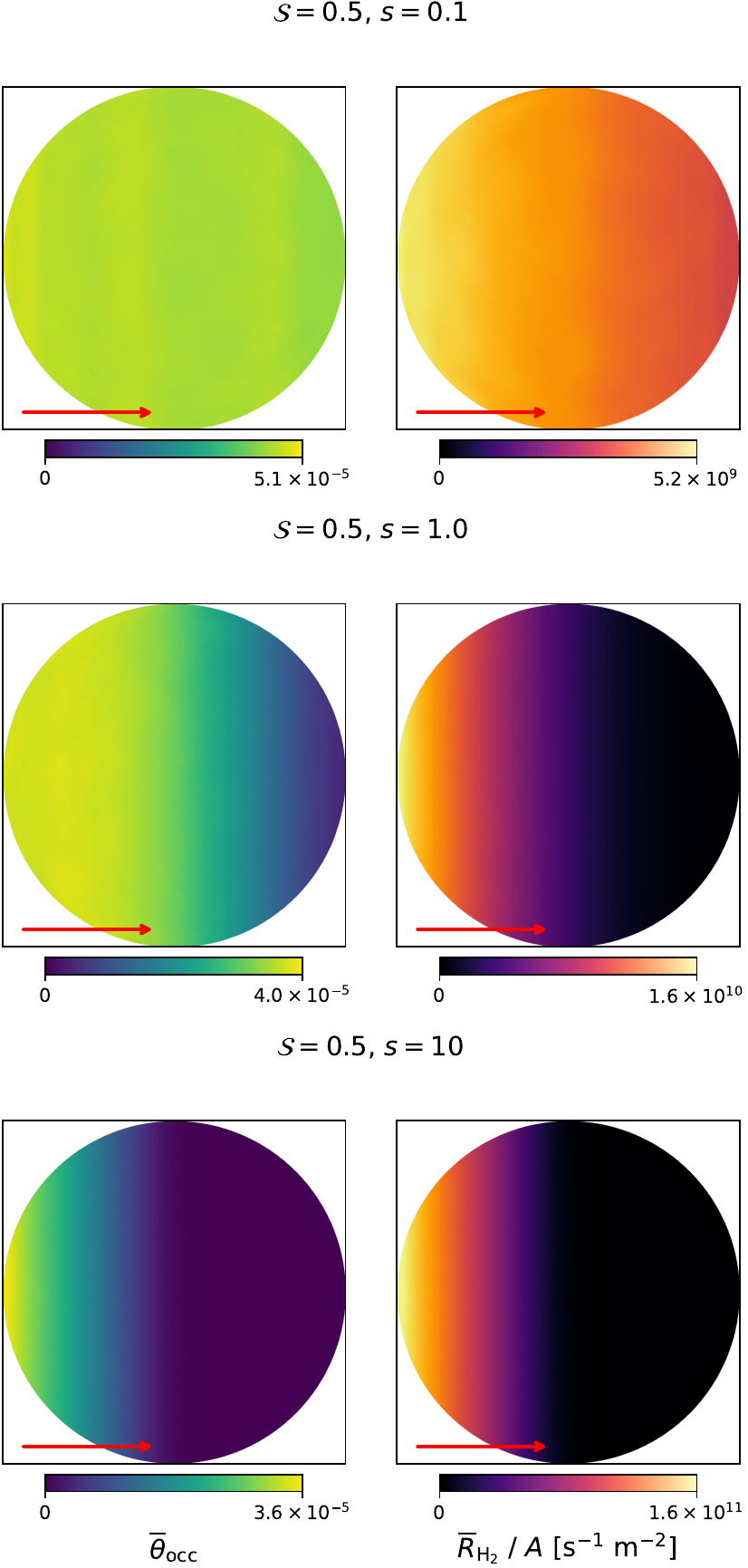}
	\end{center}
\caption{As Fig.~\ref{fig:SurfaceVar} but with a constant sticking function $\mathcal{S} = 0.5$.}
\label{fig:SurfaceConst}
\end{figure}

\begin{figure*}[ht!]
	\begin{center}
	\includegraphics[width=1.0\textwidth]{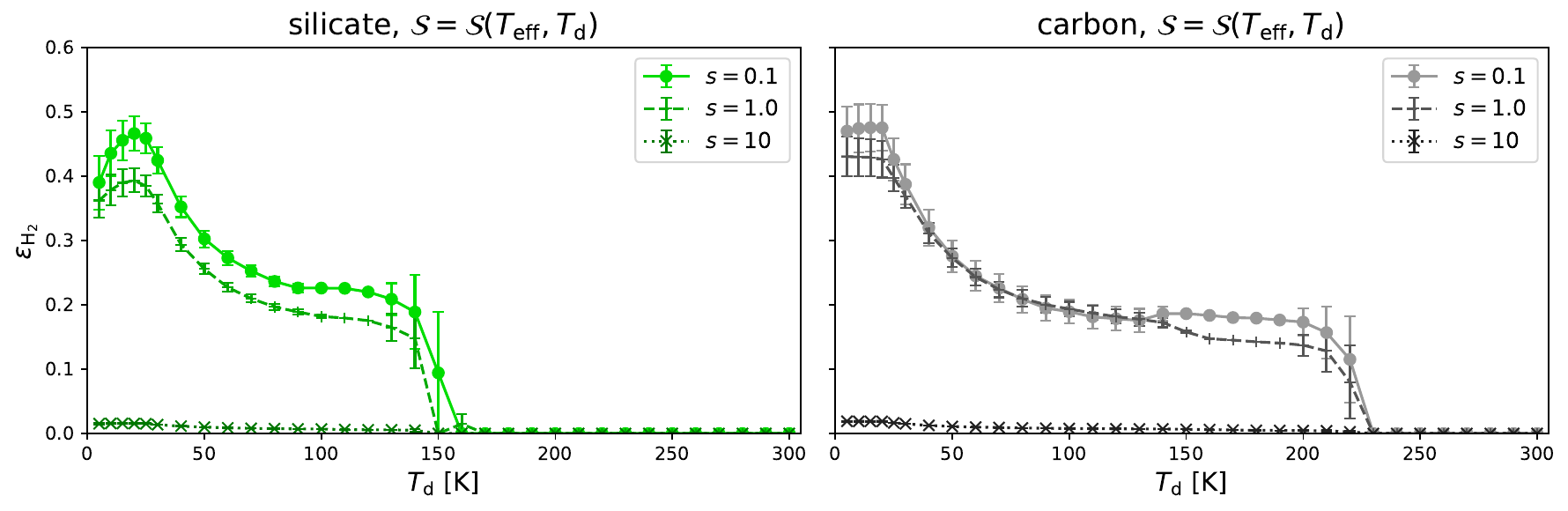}
	\end{center}
\caption{As Fig.~\ref{fig:TauVar}, but showing the H$_{2}$ formation efficiency $\epsilon_{\mathrm{H_2}}$. }
\label{fig:EffVar}
\end{figure*}

\begin{figure*}[ht!]
	\begin{center}
	\includegraphics[width=1.0\textwidth]{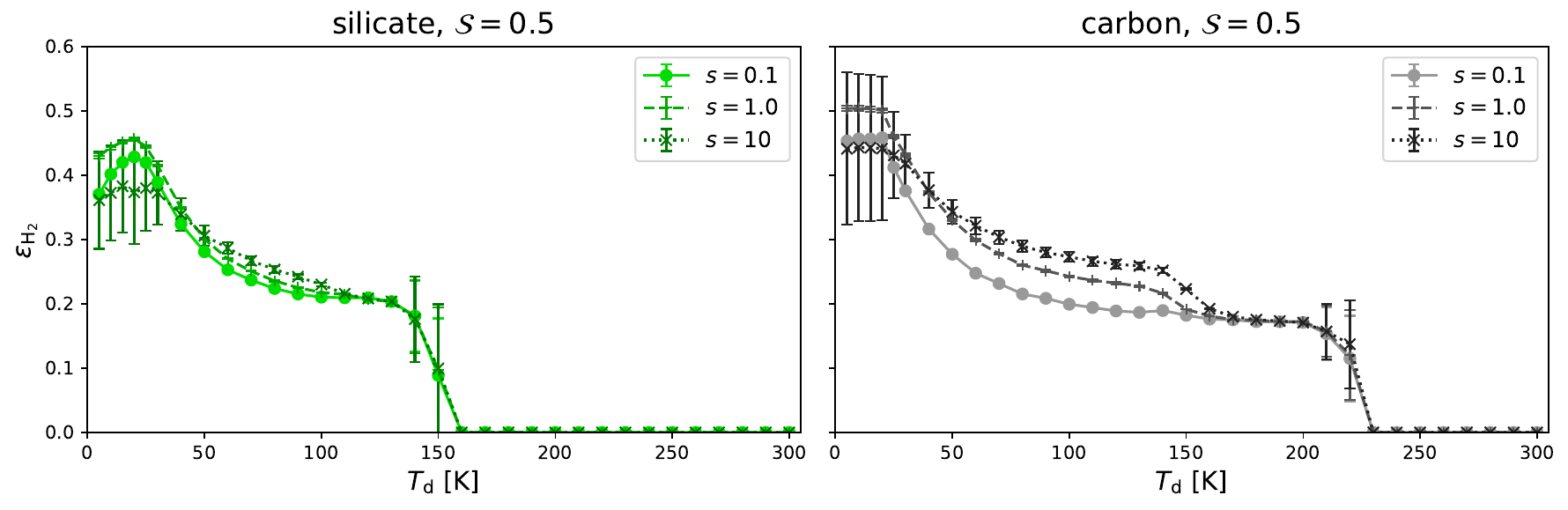}
	\end{center}
\caption{As Fig.~\ref{fig:EffVar} but for a constant sticking function $ \mathcal{S} = 0.5$. }
\label{fig:EffConst}
\end{figure*}

\section{Results}
\label{sect:Results}
KMC simulations are performed for the default input parameters listed in Table~\ref{tab:fiducial} with silicate and carbon grain materials for a variety of drift parameters, sticking functions, and dust temperatures. In addition to the equilibrium condition (see Sect.~\ref{sect:KMCscheme}), we demand at least a few thousand H$_2$ formation events for a run to complete. Runs are repeated for each permutation of input parameters with $32$ different random seeds in order to minimize and quantify the range of the KMC noise. 

\subsection{Occupation timescale}
\label{sect:ResTau}
In Fig.~\ref{fig:TauVar} we show the characteristic timescale $\tau_{\mathrm{occ}}$ it takes for the surface population on a grain to reach equilibrium. For silicate grains, H$_2$ formation breaks down at a dust temperature $T_{\mathrm{d}} = 110\ \mathrm{K}$ as thermal desorption of chemisorbed H atoms becomes dominant (compare Fig.~\ref{fig:MigrationRates}). Hence, this threshold is a direct result of the surface reactions and independent of the drift parameter $s$.

For silicate grains, the occupation timescale depends strongly on both
$T_{\mathrm{d}}$ and the drift parameter. Using the variable sticking
function $\mathcal{S}(T_{\mathrm{eff}},T_{\mathrm{d}})$,
$\tau_{\mathrm{occ}}$ reaches values of up to $\sim17\ \mathrm{yr}$ for
$s=0.1$. For strong drift, $s=10$, the occupation timescale is generally
longer over most of the temperature range. This reflects the reduced
sticking probability at high drift velocities, which slows the build-up
of the surface population and hence the approach to equilibrium.

For carbonaceous grains, H$_2$ formation remains efficient up to
$T_{\mathrm{d}}=150\ \mathrm{K}$. The occupation timescale shows less
variation with $T_{\mathrm{d}}$ than for silicate grains, with typical
values of $\tau_{\mathrm{occ}}\simeq10$--$30\ \mathrm{yr}$. This weaker
variation is associated with the more efficient p--c transition, which
allows chemisorption sites to be populated more rapidly and helps maintain
the surface population over a broader range of dust temperatures.

With a constant sticking function of $\mathcal{S} = 0.5$ as shown in Fig.~\ref{fig:TauConst}, the result for a drift factor of $s=0.1$ is comparable to the results with a temperature-dependent sticking function. However, with an increase in $s$, the  
rate of H atoms $R_\mathrm{coll}$ colliding
with the grain surface increases as well. Since the increase in $R_\mathrm{coll}$ is not compensated in this case by decreasing sticking, the occupation timescales $\tau_{\mathrm{occ}}$ go down to 1--3\ yr for silicates while for carbon grains they can reach equilibrium in less than a year for a drift factor of $s=10$.

It follows from these results that any newly formed dust grain starts to form H$_2$ molecules almost immediately compared to astronomical and cosmological timescales, provided that the local environmental conditions are favorable for H$_2$ formation. 

\subsection{Spatial distribution of surface occupation }
The advantage of KMC modeling is that the grain surface is actually spherical, so gas trajectories including drift as well as surface chemistry can be followed in 3D space. Hence, the effect of gas--dust drift can be analyzed by looking at the spatial distribution of the surface occupation and H$_2$ formation rate on the grain surface itself. 

Fig.~\ref{fig:SurfaceVar} shows the results when the sticking function ${ \mathcal{S}=\mathcal{S}(T_{\rm eff},T_{\rm d}) }$ depends on the effective gas temperature $T_{\rm eff}$, whereas Fig.~\ref{fig:SurfaceConst} shows the corresponding distributions with constant sticking probability $\mathcal{S}=0.5$. 
Silicate grains show comparable results but with slightly lower formation rates and, hence, we focus only on carbon grains in the following.

At low drift velocity, with drift parameter $s=0.1$, both the surface occupation fraction $\overline{\theta}_{\rm occ}$ and the H$_2$ formation
rate per area $A$ associated with a Fibonacci point,  $R_{\rm H_2}/A$ show only moderate spatial variations across the grain surface. In particular, $R_{\rm H_2}/A$ varies by about $30\ \%$, indicating that a weak drift-induced anisotropy is already present. Nevertheless, the thermal component of the gas velocity remains dominant over the systematic drift, and the contrast between the upstream and
downstream hemispheres is therefore considerably weaker than at larger values of $s$.

As the drift velocity increases, the surface distribution becomes
progressively more anisotropic. For $s=1$, both the occupation fraction and the H$_2$
formation rate are enhanced on the upstream hemisphere, whereas the
downstream hemisphere is depleted. This asymmetry is a
direct consequence of the directional gas flow in the grain rest frame. The spatial distribution of H$_2$ formation closely follows this anisotropic bombardment, with most reactions occurring on the hemisphere exposed to the incoming gas.

For strong drift with $s=10$, the gas flow becomes highly directional and the difference between the two sticking prescriptions becomes particularly
pronounced. In case of a sticking function $S=S(T_{\rm eff},{\rm d})$ at $s=10$, the occupation is no longer largest at the point of strongest direct bombardment. Instead, a relatively narrow band of enhanced occupation appears away from the most strongly bombarded upstream region. This behavior results from the competition between the
increasing incident collision rate and the decreasing sticking probability. At the leading edge of the sphere, where the rate $R_\mathrm{coll}$ is maximal, the component of the drift velocity $\vec{U}$ normal to the grain
surface is largest, resulting in a larger $T_{\rm eff}$ and hence a lower
probability for incident H atoms to be adsorbed. Toward the downstream regions of the surface, the drift component decreases, allowing a larger fraction of incident atoms to remain adsorbed. However, at the same time, the rate of atoms striking the grain decreases as the angle between the surface normal and the drift velocity increases, reaching a minimum once the surface normal and $\vec{U}$ are anti-parallel. The resulting maximum in surface occupation and related H$_2$ formation therefore occurs in a ring-like region perpendicular to the direction of $\vec{U}$ where the combination of incident collision rate and sticking probability is most favorable.

A qualitatively different occupation pattern is obtained when the sticking is constant. In this case, the surface occupation is concentrated toward the upstream side of the grain, and the H$_2$ formation rate reaches its largest value in the same region. The increased collision rate $R_\mathrm{coll}$ at high drift velocity therefore produces a strong enhancement of the local H$_2$ formation rate on the upstream hemisphere.

These spatial distributions demonstrate that gas--dust drift introduces a
strong directional dependence into grain-surface chemistry even for a
spherical grain. For a constant sticking probability, increasing drift
primarily enhances the incident collision rate $R_\mathrm{coll}$  and therefore the H$_2$ formation rate on the leading hemisphere. Once the velocity dependence of sticking is included, however, the increased collision rate competes with the reduced probability of adsorption at large impact velocities. At high drift velocities this competition substantially suppresses H$_2$ formation and shifts the region of highest surface occupation away from the point of maximum gas bombardment.

\subsection{The H$_2$ formation efficiency}
In Fig.~\ref{fig:EffVar} we show the formation efficiency $\epsilon_{\mathrm{H_2}}$ for silicate and carbon grains as a function of dust temperature $T_{\mathrm{d}}$ and a sticking function $\mathcal{S}(T_{\rm eff}, T_{\rm d})$. For an increasing drift $s$, the reduction of the sticking
probability substantially changes $\epsilon_{\mathrm{H_2}}$. For silicate grains and a moderate drift with $s=0.1$--1 the efficiency starts at about $\epsilon_{\mathrm{H_2}}=0.28$--0.48 in the low dust temperature regime but declines for $T_{\rm d} > 20\ \mathrm{K}$, reaching values of $\epsilon_{\mathrm{H_2}}=0.14$--0.24 at $T_{\mathrm{d}} = 140\ \mathrm{K}$. At higher values of $T_{\mathrm{d}}$, the H$_{2}$ formation efficiency rapidly drops to zero. For strong drift, on the other hand, H$_{2}$ formation is strongly suppressed at all dust temperatures, with 
$\epsilon_{\mathrm{H_2}} < 0.03$ over the entire dust temperature range.
Similarly, for carbon grains we find that for low or moderate drift parameters, $\epsilon_{\mathrm{H_2}}=0.45$--0.48 up to $T_{\mathrm{d}} = 240\ \mathrm{K}$, above which H$_{2}$ formation rapidly ceases, but for a drift parameter $s = 10$, $\epsilon_{\mathrm{H_2}} < 0.04$ for all dust temperatures. 

As discussed in Sect.~\ref{sect:ResTau}, this difference between the materials reflects the stronger binding of H atoms to the c-sites of the carbon surface. Consequently, c-H atoms remain on carbon grains up to higher dust temperatures $T_{\mathrm{d}}$, maintaining a sufficiently large surface population for H$_2$ formation. On silicate grains, the weaker binding leads to a more rapid depletion of the surface population with increasing $T_{\mathrm{d}}$, causing the H$_2$ formation efficiency to decline at lower temperatures.

Fig.~\ref{fig:EffConst} shows the corresponding results when the sticking probability is assumed to be constant with $S=0.5$. For both grain materials, the formation efficiency decreases with increasing dust temperature $T_{\mathrm{d}}$. At intermediate and high temperatures of $T_{\mathrm{d}}>25\ \mathrm{K}$, increasing the gas--dust drift enhances the H$_2$ formation efficiency $\epsilon_{\mathrm{H_2}}$. 
This enhancement results from the increased rate of H atoms colliding with the grain at larger drift velocities. In the low-temperature regime, however, $\epsilon_{\mathrm{H_2}}$ for $s=10$ remains below that for $s=1$. For a constant sticking probability of $S=0.5$, this reduction arises primarily from the directional nature of the drift, which restricts the fraction of the grain surface that is efficiently populated by incoming H atoms. At $T_{\mathrm{d}} \gtrsim 25\ \mathrm{K}$, this effect is compensated by the increased H-atom rate $R_\mathrm{coll}$, resulting in an enhanced formation efficiency. For the variable sticking model, strong drift additionally reduces the sticking probability because of the increased impact energy, further decreasing the effective surface area available for H$_2$ formation
(compare Figs.~\ref{fig:SurfaceVar} and \ref{fig:SurfaceConst}).

Overall, the efficiency rises up to about $\epsilon_{\mathrm{H_2}}=0.48$ for both silicate and carbon in the low temperature regime and up to $\epsilon_{\mathrm{H_2}}=0.25$ and $\epsilon_{\mathrm{H_2}}=0.15$ for silicate and carbon, respectively, in the high temperature regime located shortly before H$_2$ formation ceases.
The larger collision rate at high drift velocity $U$ is therefore offset by the reduction in the probability that an incident H atom sticks to the grain. Consequently, the enhancement of H$_2$ formation with drift seen for constant $S$ does not persist when the sticking probability is allowed to respond to the increased impact energy.

We note that increasing the gas number density $n_{\rm H}$ increases both the collision rate and the H$_2$ formation rate. The resulting formation
efficiency is therefore expected to depend more weakly on density, although changes in the surface population can introduce an additional density dependence  \citep[see, e.g.][]{Iqbal2012,Das2026arXiv}.
At higher gas temperatures, the overall sticking probability decreases, which tends to reduce $\epsilon_{\rm H_2}$.

\begin{figure*}[ht!]
	\begin{center}
	\includegraphics[width=1.0\textwidth]{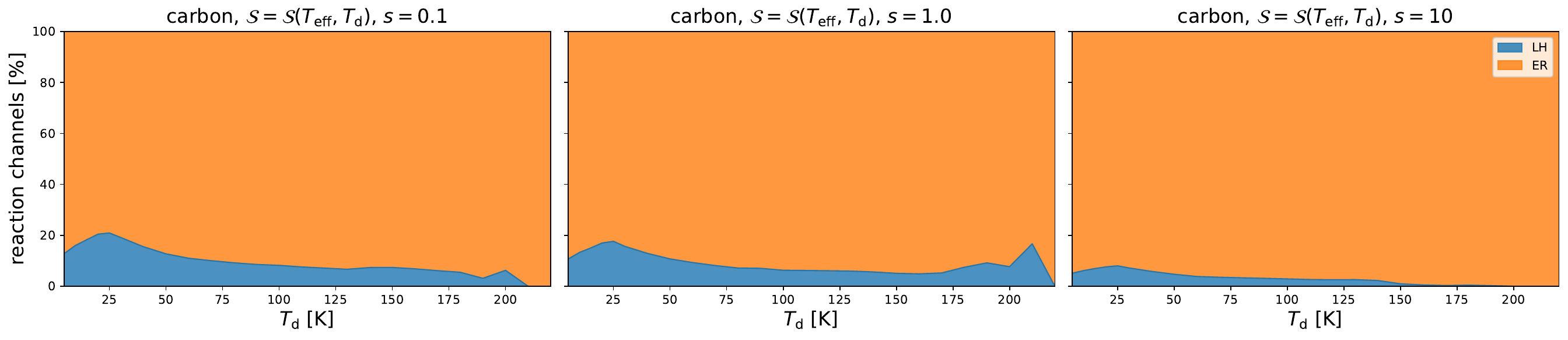}
	\end{center}
\caption{ The fraction of H$_2$ molecules formed by the LH (blue area) and the ER (orange area) mechanisms, plotted as a function of the dust temperature $T_{\mathrm{d}}$ for a drift parameter of $s=0.1$ (left panel), $s=1.0$ (middle panel), and $s=10$ (right panel), and a temperature-dependent sticking function $\mathcal{S} = \mathcal{S}( T_{\mathrm{eff}} , T_{\mathrm{d}})$.}
\label{fig:ChannelVar}
\end{figure*}

\begin{figure*}[ht!]
	\begin{center}
	\includegraphics[width=1.0\textwidth]{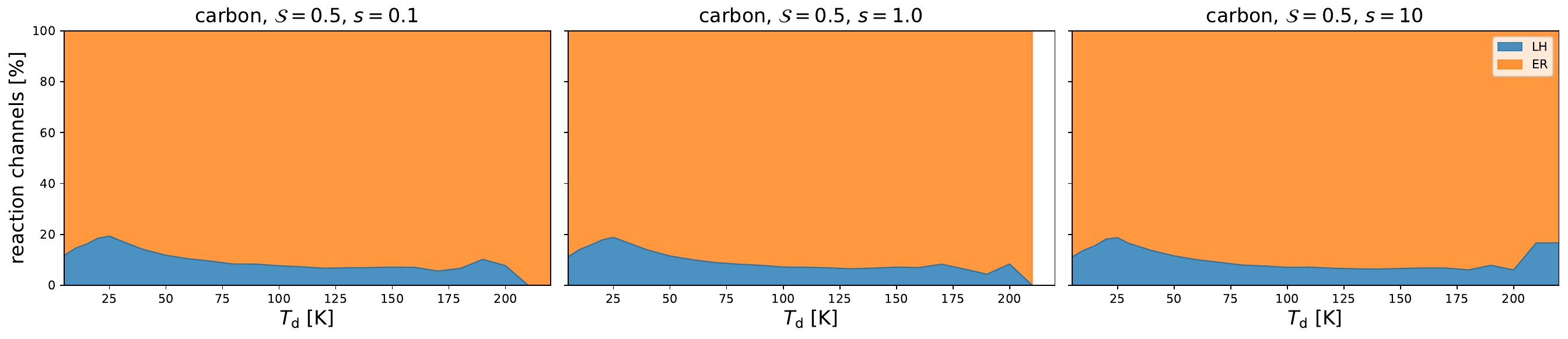}
	\end{center}
\caption{The same as Fig.~\ref{fig:ChannelVar} but for a constant sticking function $ \mathcal{S} = 0.5$.}
\label{fig:ChannelConst}
\end{figure*}

\subsection{Reaction channels}
In Figs.~\ref{fig:ChannelVar} and~\ref{fig:ChannelConst} we show the relative contributions of the ER and LH channels for carbon grains as a function of dust temperature $T_{\mathrm{d}}$ and drift parameter using either a drift-dependent or a constant sticking function. The ER mechanism dominates H$_2$ formation over the entire parameter range considered. At low and moderate drift velocities $s=0.1$ and $1$, the relative contributions show a modest LH contribution at low temperatures of up to 20\%, decreasing further toward higher $T_d$. For the highest temperature before H$_2$ formation breaks down in the low and moderate drift cases, the LH mechanism increases again slightly as increasing cc mobility supports more efficient formation on the grain surface.  


This behavior reflects the sensitivity of the LH mechanism to the surface population. When the sticking probability  is drift-dependent, the larger effective impact energy at high drift velocities reduces the probability that incoming H atoms are adsorbed. This lowers the population available for surface migration and therefore suppresses the probability of an encounter between two adsorbed H atoms. The LH channel is consequently reduced more strongly than the ER channel. The latter can still proceed whenever an incoming gas-phase H atom encounters an H atom already present on the surface.


The relative fractions of ER and LH formation obtained for silicate grains are rather similar to those reported for carbonaceous grains. We therefore presented only the carbon results in this section.

\section{Discussion}
\label{sect:Discussion}
Previous studies have already shown that H$_2$ formation can remain efficient under conditions where simple low-temperature LH models fail. \citet{LeBourlot2012} demonstrated that ER reactions can maintain substantial H$_2$ formation in warm photodissociation regions (PDRs), while \citet{Bron2014} showed that stochastic grain-temperature fluctuations can modify the formation rate, particularly for small grains. Laboratory experiments further demonstrate efficient H$_2$ formation on carbonaceous surfaces up to $T_{\rm d}\sim250\ \mathrm{K}$ \citep{Grieco2023}. Consistent with these studies, we find that carbonaceous grains sustain H$_2$ formation to considerably higher temperatures than silicate grains. We further find that the modest LH contribution is largest at low temperatures, whereas ER becomes increasingly dominant toward higher $T_{\rm d}$, in agreement with previous microscopic models
\citep{Iqbal2012,Das2026arXiv}. This transition results from the decreasing residence time of p-H atoms because thermal desorption suppresses the surface population required for LH reactions, while ER reactions can proceed through direct collisions of gas-phase H with adsorbed H.

Our results do show, however, that systematic gas--dust drift can substantially modify this picture. At moderate drift, the increased H-atom collision rate can enhance H$_2$ production, whereas at large drift velocities the increased impact energy reduces the sticking probability and can strongly suppress the
formation efficiency. 

This dependence on drift is potentially important for interpreting H$_2$ observations. H$_2$ formation affects the H/H$_2$ transition, gas heating, and rotational and rovibrational emission, which are widely used to constrain the physical conditions of PDRs and shocked gas. \citet{Wakelam2017} emphasized the diagnostic potential of spatially resolved H$_2$ emission,
including in turbulent environments with large velocity dispersions. Our results suggest that variations in inferred H$_2$ formation rates may
therefore partly reflect gas--dust drift rather than variations in density, radiation field, grain temperature, or composition alone. In quiescent cool atomic gas, grains are generally expected to remain sufficiently coupled to the gas that
$s\lesssim1$, making strongly superthermal drift comparatively uncommon. Values $s>1$ may, however, arise locally in dynamically active environments.
Molecular outflows typically reach velocities of order $10~{\rm km\,s^{-1}}$, while protostellar jets can exceed $100~{\rm km\,s^{-1}}$ \citep[e.g.][]{Bally2016}. Although these bulk velocities do not directly correspond to gas--dust drift velocities, substantial grain--gas decoupling can occur in shocks. In C-type shocks, grain--neutral drift velocities of order $10~{\rm km\,s^{-1}}$ are possible \citep[e.g.][]{Guillet2011,VanLoo2013}, corresponding to $s\simeq7$ at $T_{\rm g}=100$ K and $s\simeq2$ at $T_{\rm g}=1000$ K. Other possible
drivers of superthermal drift include cloud--cloud collisions, stellar winds, and radiation pressure \citep{Habing1994,Schirmer2022}. Moreover, MHD turbulence can accelerate charged
grains through interactions with turbulent magnetic fluctuations, with gyroresonant acceleration by fast MHD modes predicted to produce supersonic
grain velocities in the cold neutral medium
\citep{YanLazarian2003,Yan2004}. Thus, while $s\lesssim1$ is likely most representative of quiescent atomic gas, the $s>1$ regime explored here may be relevant locally in shocks, colliding flows, and strongly turbulent or irradiated environments. We therefore regard $s=10$ primarily as a limiting case representing extreme gas--dust decoupling rather than typical diffuse-ISM conditions.

Beyond chemistry, drift-dependent H$_2$ formation may also affect grain dynamics. The ejection of newly formed H$_2$ may transfer energies of order $0.2\ \mathrm{eV}$ \citep{Hunter1978} onto the grain surface and produce systematic torques \citep{Purcell1979,DraineWeingartner1997}.  \citet{Hoang2009ApJ} showed that such torques can modify grain rotation and enhance alignment in
combination with radiative torques. Because the H$_2$ torque depends on the incident H collision rate and formation efficiency, our results imply that its
strength may itself depend on gas--dust drift  as increasing drift enhances the incident rate but can simultaneously suppress sticking and H$_2$ formation.

A possible observational test is provided by the IC~63 PDR, where ultraviolet radiation from the star $\gamma$ Cassiopeiae bombards the gas and dust in its vicinity. Here,  \citet{Andersson2013ApJ} found enhanced optical polarization toward regions
of strong H$_2$ fluorescence, providing evidence for H$_2$-formation-driven grain alignment. Due to IC~63's intense radiation field and dynamically structured PDR, significant gas--dust drift of order $0.1$--$0.3~{\rm km\,s^{-1}}$ may occur in this region \citep[e.g.][]{Schirmer2022}, although its magnitude has not been directly constrained observationally. The gas temperature varies considerably throughout IC~63, with temperatures of order $T_{\rm g}\sim10$--$100~{\rm K}$ characteristic of the cooler molecular material and a column-density-averaged temperature of approximately $50~{\rm K}$ \citep{Jansen1994,Andersson2013ApJ}. For this temperature range, drift velocities of $0.1$--$0.3~{\rm km\,s^{-1}}$ correspond to a drift parameter of $s\simeq0.07$--$0.65$. PDRs such as IC~63 may therefore provide an opportunity to test whether drift-dependent sticking and H$_2$ formation
translate into measurable changes in grain alignment and the resulting dust-polarization signal.

Finally, we emphasize that a major remaining uncertainty in our work is the
lack of a well-constrained sticking function over the relevant range of gas
temperatures and drift velocities. As recent studies suggest, a promising way
forward is to determine the sticking probability systematically as a function
of impact energy and grain material using first-principles molecular dynamics
simulations  \citep{Goumans2009,Morrissey2022Icar,Morrissey2026AJ}. Moreover, both physisorption and chemisorption may depend on the pre-existing surface coverage, since already adsorbed species can modify the local interaction potential and available adsorption sites. Molecular dynamics simulations may help to quantify such coverage-dependent effects and determine how they alter the probability of physisorption and chemisorption. Such simulations could therefore provide physically motivated sticking functions while simultaneously placing further constraints on the rates of relevant surface processes, thereby reducing some of the key uncertainties in future H$_2$ formation models.





\section{Summary and Outlook}
\label{sect:Summary}
In this work, we have presented a newly developed event-driven KMC model for H$_2$ formation on spherical silicate and carbonaceous dust grains. The model follows individual H atoms from their arrival at the grain through adsorption,
surface migration, desorption, and H$_2$ formation via the LH and ER mechanisms. In contrast to previous microscopic models, we explicitly include systematic gas--dust drift by sampling the incident atoms from a shifted Maxwellian velocity distribution. This allows both the enhancement to the collision rates and the anisotropic bombardment of the grain surface to be treated consistently. We further compare constant sticking with a drift-dependent sticking prescription in which the increased impact energy modifies the local adsorption probability.

The main findings of this work are:
\begin{itemize}

\item gas--dust drift introduces a strong anisotropy in the surface occupation and H$_2$ formation rate, even for spherical grains. At large drift velocities, H$_2$ formation becomes concentrated in a ring perpendicular to the velocity vector.

\item For a constant sticking probability of $\mathcal{S}=0.5$, increasing drift generally enhances H$_2$ formation because of the increased rate of H atoms colliding with the grain. However, this enhancement disappears when the effect of the increased impact energy on sticking is included.

\item With drift-dependent sticking, strong drift substantially suppresses H$_2$ formation 
efficiency to  $\epsilon_{\mathrm{H_2}} < 0.03$ at drift velocity $U= 10 v_\mathrm{th}$ compared with efficiencies of 20--50\% at low and moderate drift.

\item Carbonaceous grains maintain efficient H$_2$ formation to considerably higher dust temperatures than silicate grains because of their stronger chemisorption binding. Nevertheless, strong drift suppresses the efficency for both materials.

\item ER reactions dominate H$_2$ formation across the investigated parameter space. Strong drift further suppresses the LH contribution because reduced sticking lowers the surface population available for diffusion and subsequent LH reactions.

\item Consequently, the H$_2$ formation rate cannot in general be determined from gas and dust temperatures alone. In dynamical environments, the gas--dust relative velocity and its effect on sticking constitute additional parameters that should be considered in H$_2$ formation prescriptions.

\end{itemize}
The present model provides a first step toward coupling drift-dependent surface chemistry to the dynamics of interstellar grains. This model can be applied to rotating and irregular grains
to investigate how grain morphology, shadowing, gas--dust drift, surface chemistry, and grain rotation jointly determine H$_2$ formation and the dynamical evolution of realistic interstellar dust.

\begin{acknowledgements}
S.R. and R.S.K. acknowledge financial support from the Heidelberg cluster of excellence (EXC 2181 - 390900948) “{\em STRUCTURES}: A unifying approach to emergent phenomena in the physical world, mathematics, and complex data”, specifically via the exploratory project EP 4.4 "On the Alignment Dynamics of Interstellar Dust" as well as funding form the European Research Council in the ERC synergy grant “{\em ECOGAL} – Understanding our Galactic ecosystem: From the disk of the Milky Way to the formation sites of stars and planets” (project ID 855130). The project made use of computing resources provided by {\em The L\"{a}nd} through bwHPC and by DFG through grant INST 35/1134-1 FUGG. Data are in part stored at SDS@hd supported by the Ministry of Science, Research and the Arts and by DFG through grant INST 35/1314-1 FUGG. M.-M.M.L. was partly supported by NSF grant AST23-07950.
\end{acknowledgements}

\bibliographystyle{aa}
\bibliography{./bibtex}

\end{document}